\documentstyle[pre,aps]{revtex}                       
\def\bfr {{\bf r}}
\def\bfrr {{\bf R}}

\def\bfk {{\bf k}}

\begin{document}
\draft
\title{Statistical Properties of the Dense Hydrogen Plasma: an ab initio 
Molecular Dynamics Investigation}
\author{Jorge Kohanoff$^{~\# *}$ and Jean-Pierre Hansen$^*$}
\address{
$^{\#}$ International Centre for Theoretical Physics, Strada Costiera 11, 
I-34014 Trieste, Italy} 
\address{
$^*$ Laboratoire de Physique (U.R.A. 1325 du CNRS), Ecole Normale 
Sup\'{e}rieure de Lyon, F-69364 Lyon Cedex 07, France}
\date{\today} 
\maketitle
\begin{abstract}
The hydrogen plasma is studied in the very high density (atomic and metallic)
regime by extensive ab initio Molecular Dynamics simulations. Protons are
treated classically, and electrons in the Born- Oppenheimer framework,
within the local density approximation (LDA) to density functional theory.
Densities and temperatures studied fall within the strong coupling regime
of the protons. We address the question of the validity of linear screening,
and we find it to yield a reasonably good description up to $r_s\approx 0.5$, 
but already too crude for $r_s=1$ (with $r_s=(3/4\pi\rho)^{1/3}$ the ion sphere
radius). These values are typical of Jovian planets interiors. Finite-size 
and Brillouin zone sampling effects in metallic systems are studied 
and shown to be very delicate also in the fluid (liquid metal) phase. We 
analyse the low-temperature phase diagram and the melting transition. A
remarkably fast decrease of the melting temperature with decreasing density 
is found, up to a point when it becomes comparable to the Fermi temperature 
of the protons. The possible vicinity of a triple point bcc-hcp(fcc)-liquid 
is discussed in the region of $r_s\approx 1.1$ and $T\approx 100~-~200~K$.
The fluid phase is studied in detail for several temperatures. The
structure of the fluid is found to be reminiscent of the underlying bcc
(solid) phase. Proton-electron correlations show a weak temperature 
dependence, and proton-proton correlations exhibit a well-defined first
coordination shell, thus characterizing fluid H in this regime as an 
atomic liquid. Diffusion coefficients are computed and compared to the values 
for the one-component plasma (OCP). Vibrational densities of states (VDOS) 
show a plasmon renormalization due to electron screening, and the presence of 
a plasmon-coupled single-particle mode up to very high temperatures. 
Collective modes are studied through dynamical structure factors. In 
close relationship with the VDOS, the simulations reveal the remarkable 
persistence of a weakly damped high-frequency ion acoustic mode, even 
under conditions of strong electron screening. The possibility of using 
this observation as a diagnostic for the plasma phase transition to the fluid 
molecular phase at lower densities is discussed.
\end{abstract} 

%\narrowtext
\twocolumn

\section{Introduction and state of the art}

The possibility of hydrogen metallization under high pressure 
was first discussed by Wigner and Huntington in the thirties~\cite{wigner}.
This particular subject was included in the more general 
conjecture that any system becomes metallic if sufficient pressure is 
applied. Electrons, which are bound in the isolated atomic or molecular 
species, hybridize in a condensed phase to form extended states which gather 
in energy bands. When pressure is applied, the bands widen due to the 
enhancement of the overlap between neighboring atoms' orbitals, up to 
the point at which the energy gap between the valence and the conduction 
band vanishes, thus giving rise to metallic behavior. This enhancement of 
the overlap can also be viewed as the growing importance of the electronic 
kinetic energy relative to the ion-electron potential; the latter tends 
to bind electrons to the ions (also in the form of interatomic, 
intramolecular or intermolecular bonds). In the case of pure hydrogen,
metallization was estimated to occur around 2.5 Mbars (1 Mbar = 100 GPa), 
i.e. at pressures that have become accessible to experiment in diamond anvil 
cells only very recently~\cite{mao}. The absence of clear signs of metallic 
behaviour up to 2.9 Mbar~\cite{cui,ruoff} certainly adds to the
challenge, and stimulates the continuous improvement of experimental 
setups in the search for the {\it reluctant} metallization. 

An additional difficulty in the description of this particular 
metal-insulator phase transition arises from the fact that at low 
pressures the low-temperature phase is not a monoatomic but a molecular 
solid, i.e. an hexagonal close packed (hcp) arrangement of H$_2$ 
molecules~\cite{mao}. In this sense, H behaves more like a halogen than 
like an alkali metal~\cite{ashcroft}. Up to pressures of the order of 1 
Mbar, experiment indicates that the hcp arrangement is preserved. 
At pressures of the order of 110 GPa, the low-pressure free-rotator phase 
of para-H$_2$ freezes into an orientationally ordered phase, whose nature 
is not yet fully understood. The observation of more than one vibron 
mode~\cite{hanfland} precludes the hcp structure with all the 
molecules pointing along the c-axis. However, other more complex hcp 
structures with tilted molecules (herringbone-like) are compatible with 
experiment~\cite{mao}. It is also possible that orientational order is
not complete, thus giving rise to a sequence of weak phase transitions 
until the perfectly ordered phase is reached. 

An intriguing phase transition at 150 GPa, signalled by a discontinuity in 
the molecular vibron~\cite{hemley}, has been ascribed either to a relative 
reorientation of the H$_2$ groups~\cite{kaxiras}, to metallization arising 
from an electronic band-overlap mechanism without molecular 
reorientation~\cite{chacham}, and to both occuring at the same 
time~\cite{surh}. Density functional 
(DFT) calculations indicate that arrangements with complex molecular 
orientations (non parallel) are energetically preferred in the high 
pressure phase~\cite{kaxiras}. Another study~\cite{nagara} concentrated 
on other classes of structures including cubic $Pa3$ and rutile, and also 
two hcp structures, namely $Pca2_1$ and $P2_1/c$. The hexagonal $Pca2_1$
was found to be the favorable one above the 150 GPa transition. Very
recently, Tse and Klug performed ab initio simulated annealing calculations
with 96 H atoms in the supercell, and found as a ground state an orthorombic 
structure, instead of hcp, composed of groups of three strongly interacting 
$H_2$ molecules~\cite{tse}. All these candidates were found to be insulating,
and the reason for this was claimed to be the opening of a hybridization
gap at the Fermi level~\cite{mazin}.
Diffusion Monte Carlo (DMC) calculations by Natoli et al.~\cite{natoli1} 
have essentially confirmed results by Kaxiras et al.~\cite{kaxiras}, also 
at the quantitative level for the angles between different molecular units. 
They  considered neither Nagara's $Pca2_1$ nor Tse's orthorombic 
structure. This DMC calculation also finds an insulating behaviour and, 
interestingly, zero-point-motion effects due to the protons appear to be 
quite structure independent. As a consequence, zero-point-motion 
would turn out to be irrelevant as regards the 150 GPa 
transition, contrarily to the claims  of Surh et al.~\cite{surh}.

Well-converged local density functional (LDA) calculations supplemented 
with the zero-point-motion energy contribution taken from frozen-phonon 
calculations~\cite{barbee}, suggest that the molecular-hcp phase goes over 
directly to an atomic squeezed-hexagonal phase, by breaking the 
intramolecular bonds, at a pressure of 380 GPa. However, this
calculation did not take into account non-hexagonal structures. Hence, 
also from the theoretical point of view, it is not clear up to now whether 
metallization happens already in the molecular
phase, or whether it is a consequence of dissociation, thus occuring 
simultaneously with the molecular-atomic transition~\cite{cui}. A 
rhombohedral structure was also postulated to supersede the 
squeezed-hexagonal phase at higher pressures (above 400 GPa)~\cite{barcoh}, 
while the transition to a body-centered-cubic (bcc) phase was located at 
around 1100 GPa~\cite{barbee}. 

In general, static methods suffer from the drawback of having to carry out  
the investigation by {\it guessing} 
different structures. In this respect, the molecular dynamics results
by Tse and Klug seem to be the most reliable. However, there is still
some doubt because it is not obvious that their 96-atom supercell with 
$\Gamma$-point sampling and fixed cell volume and shape calculation is 
a sufficiently good description. A recently proposed method that combines 
state-of-the-art electronic structure calculations with variable cell 
shape Molecular Dynamics simulations~\cite{CPPR}, thus avoiding every 
undesired bias, is currently being used to investigate the low-temperature 
phase of H as a funcion of pressure~\cite{ksct:unp}.

It is important to remark that, in general, energy differences between
different structures are quite small, so that different levels of 
approximation often lead to different ground-state structures. In this
respect, fully quantum-mechanical DMC calculations are very indicative. 
Ceperley and Alder~\cite{cepald87} located the molecular-to-atomic phase 
transiton at a pressure of about 300 GPa, by studying only cubic structures. 
Further refinements of these DMC calculations have shown how dramatic the 
effect of improving the description can be~\cite{natoli2}. Eventually, it 
appears that the atomic ground state structure in the vicinity of the 
molecular-atomic transition is the diamond structure, but the energy 
difference with respect to others is very small and, in particular, with 
respect to $\beta$-Sn and hexagonal diamond, it is well within 
the error bars. The diamond structure was considered as a candidate for 
the ground state in one of the previous DFT calculations, but found to be
less favorable than a distorted hexagonal phase~\cite{barbee}.

The monoatomic bcc structure is unequivocally identified as the ground 
state in the very high density limit. There, the electronic kinetic energy 
is largely dominant over the remaining contributions, such that the 
electrons behave essentially as free fermions; i.e. the electronic subsystem
decouples from the protons and becomes a rigid, homogeneous electron gas,
which acts only as a neutralizing background for the protons. This is
well-known as the one-component plasma (OCP) model, whose classical version 
was shown to crystallize into the bcc structure. This follows from a simple 
calculation of the Madelung energy, although the energy differences 
relative to other structures turn out to be rather small. In the 
classical OCP, which was thoroughly studied during the past two 
decades~\cite{baus}, the bcc structure turns out to be the stable one also 
at finite temperature and up to the melting point. The quantum version of 
the OCP at finite temperature has not yet been studied in detail.
A numerical study using the path integral Monte Carlo technique (PIMC) is 
currently under development~\cite{cep:pc}. 

At higher temperatures the protons begin to behave as classical particles. 
The degeneracy temperature for the protons can be estimated to be of the 
order of $T_d^p\approx 180~K/r_s^2$, by comparing the mean interprotonic 
distance and the thermal de Broglie wavelength. The electron degeneracy 
temperature ($T_d^e$) lies well above (by a factor $m_p=1836$) such that, at 
temperatures lower than $T_d^e\approx 326000~K/r_s^2=1~{\rm Hartree}/r_s^2$, 
electrons can be considered to follow the protons adiabatically, always 
staying in the the ground state compatible with the current protonic 
configuration (Fermi temperatures are about twice these numbers, e.g. 
$T_F^p=326~K/r_s^2$). This latter approximation allowed Hohl et 
al.~\cite{hohl} to examine the hot, dense {\it molecular} phase by means 
of ab initio Molecular Dynamics (AIMD) simulations, and also to make
some progress regarding the low-temperature structure. The AIMD study 
of the {\it atomic} phase is the subject of this and of a previous 
publication~\cite{kh:prl}. The effect of excited electronic states in the 
same regime has very recently been explored by a novel ab initio Molecular
Dynamics scheme using Mermin's density functional~\cite{ali}, and the 
fully-quantum PIMC method was applied to simulate the very 
high-temperature fluid phase~\cite{carlo}, where the number of excited 
electronic states involved becomes too large to be treated efficiently
with Mermin's functional. The low-temperature regime, where protons in 
turn become degenerate, requires different simulation techniques 
which are currently being developed~\cite{kpc:unp}.

High-temperature high-pressure measurements on Hydrogen have been 
very recently reported in shock wave experiments~\cite{weir}. 
Pressures between 1 and 2 Mbar and temperatures of some thousand $K$
are now feasible and rather controllable, and further expansion
of this range is in sight. In such experiments, Nellis et al. observed
the metallization of fluid molecular Hydrogen at $P=1.4$ Mbar and 
$T=3000~K$, a temperature which is significantly lower than that 
predicted for the plasma phase transition by Chabrier et 
al.~\cite{chabrier} on the basis of a multicomponent thermodynamic 
theory that employs an approximate equation of state. Although more 
accurate PIMC simulations have essentially confirmed the predictions 
of the theory, including the existence of a first order phase 
transition at high pressures and a somewhat lower value for the 
transition temeprature~\cite{magro}, the discrepancy with respect 
to shock wave experiments and AIMD results~\cite{aliala} still 
holds. The latter would rather indicate a continuous transition 
in which metallization and dissociation of H$_2$ molecules are 
closely related phenomena. A possible scenario would be the 
existence of two phase transitions: a continuous one, at relatively 
low temperatures, where molecular Hydrogen already metallizes and 
begins to dissociate, and a discontinuous one at higher temperatures 
where a massive dissociation occurs with a concomitant jump in 
the electrical conductivity.

Hydrogen at finite temperatures and high densites consitutes, hence, a
strongly coupled proton-electron plasma which is of great astrophysical
interest since it is, in particular, the major constituent in Jovian
planet interiors. Actually, the latter are basically H plasmas with a few 
percent admixture of He, and the statistical properties of the mixture
(mixing/demixing transition) are of great importance to account for 
experimental observations and to establish models for the evolution 
of these planets~\cite{chabrier1}. Nevertheless, the answer to many 
relevant questions in astrophysics requires a full understanding of the 
statistical properties of the pure main constituents. Hydrogen is present 
in both atomic and molecular fluid phases. These are separated by a 
boundary located at some distance from the center of the planet. The
characteristics of this boundary, e.g. precise location, width, etc. 
depend on the density and temperature profile, i.e. on the equation
of state. The matter of dissociation and metallization is particularly
relevant because the large magnetic field measured in Jupiter would
include a non neglgible component generated in the outer molecular 
fluid phase, provided that temperature is above the metallization 
threshold. To quantify this effect the important quantities are the
dissociated fraction and the electrical conductivity, which can be 
obtained from simulations in the molecular phase.

The hydrogen plasma is perhaps the most fundamental, simple many-body 
system, with {\it all} the interactions (proton-proton, proton-electron 
and electron-electron) described exactly by the {\it bare} Coulomb 
potential. In spite of this, the phase diagram appears to be surprisingly 
rich. The purpose of this article is to present a detailed study of the
very high pressure physical properties and phase diagram of this simple 
and fascinating system. We are going to deal always with atomic phases, 
the molecular phases appearing at lower densities than those studied here. 
In Section II we briefly describe some details of the simulations, and we 
address some important technical issues such as the interplay between 
pseudopotentials and basis set expansions. Section III is devoted to 
the analysis of the validity of linear response theory in its ability 
to describe proton-proton interactions in terms of an effective pair 
potential of the screened Coulomb form. In Section IV we address a 
crucial problem that arises in the ab initio simulation of metallic 
systems, in particular liquid metals, i.e. size effects and Fermi 
surface sampling. In Section V we enter directly into the properties 
of the Hydrogen plasma by describing the solid phases and the melting at 
very high pressure. Section VI is concerned with the fluid phase, which we 
characterize as atomic-like. Diffusion and vibrational properties are
presented in Section VII, while Section VIII is devoted to a thorough 
study of the collective dynamics, as a approach to the metal-insulator
transition coming from the metallic side. The behavior of the 
longitudinal collective mode is studied as a function of density and 
temperature, and proposed as a probe for the metal-insulator transition 
at finite temeprature. Finally, we conclude in Section IX. In the
remaining part of this Section we define some useful parameters and
ratios.

The plasma is made up of $n$ electrons and as many ions (protons) per
unit volume, such that the usual dimensionless density parameter is 
$r_s=a/a_B=(3/4\pi n)^{1/3}/a_B$, where $a$ is the mean interionic 
distance (ion-sphere radius) and $a_B$ is the Bohr radius. Adopting
atomic units throughout, the Fermi momentum is $k_F=(9/4\pi)^{1/3}/r_s$,
and the Thomas-Fermi screening length is $k_{TF}=(12/\pi)^{1/3}/\sqrt{r_s}$.
The dimensionless Coulomb coupling constant associated with the classical 
ions is $\Gamma=e^2/(ak_BT)$. Note that $k_BT=1/(\Gamma r_s)$, and that 
the electron degeneracy parameter is $\theta=T/T_F=2(4/9\pi)^{2/3}r_s/\Gamma$. 
Typical densities inside Jovian planets are between 1 and 10 $g/cm^3$
and temperatures are of the order of 100 to 10000 $K$, i.e. $r_s$ ranging 
from 0.5 to 1.5, and $\Gamma\approx 20-200$~\cite{baus}. Note that within 
this range $\theta\approx 0.015\ll~1$ and for the present calculations, 
whose aim is precisely to address some aspects of astrophysical 
plasmas, we can resort to the adiabatic approximation by assuming that the
electrons are always in their instantaneous ground state for any given 
ionic configuration. Higher temperatures require the relaxation of this 
hypothesis.

\section{Technical details of the simulations}

Our simulations were carried out using the standard Car-Parrinello (CP) 
AIMD scheme~\cite{carpar}. The ions were considered as classical point-like
particles, whereas the electrons were described by density functional
theory (DFT) within the local density approximation (LDA)~\cite{ks}. The 
electronic density was constructed via Kohn-Sham single-electron orbitals, 
expanded in a plane wave basis set up to a specified energy cutoff $E_{\rm 
cut}$ (see below). We have extensively studied a system consisting of 54 
H atoms contained in a simple cubic simulation box under periodic boundary 
conditions (PBC), and we have also performed simulations for larger 
supercells, containing 128 and 162 H atoms, in order to analyse
finite-size effects. The plane wave expansion was carried out around 
the $\Gamma$-point of the supercell's Brillouin zone. Finite-size and 
Brillouin zone sampling effects will be analysed later, in a forthcoming
section.

We used the exchange-correlation functional deduced from the results 
of quantum Monte Carlo calculations on the uniform electron 
gas~\cite{cepald80}, and the bare Coulomb potential for the ion-electron 
interaction. The singularity of the ion-electron Coulomb attraction
implies that there is never absolute convergence of the electronic 
quantities as the energy cutoff for plane waves is increased. In 
particular, the cusp condition is never satisfied, the density reaching 
the origin with zero slope. This is because the Fourier expansion 
of $1/r$ is also a long-range function ($4\pi/g^2$), meaning that there always 
exists a spatial scale small enough to require a representation involving
large-$g$ components. To satisfy the cusp condition, a different (localized)
basis set is needed instead of plane waves, e.g. Slater-type 
orbitals~\cite{harris}. A finite cutoff $g_{\rm cut}=\sqrt{2E_{\rm 
cut}({\rm Ry})}$ translates into the following effective potential:

\begin{equation}
v_{\rm eff}(r)= \frac{1}{r}~\left(1-\frac{2}{\pi}\int_{g_{\rm cut}}^{\infty}
\frac{\sin(x)}{x}~dx\right)
\end{equation}

This description is variational in the number of plane waves (i.e. in the 
energy cutoff) and, even if the energy never fully converges, 
other quantities, like the electronic distribution around the 
protons, are affected only in a tiny region around the nuclei, the remainder 
being essentially converged at a finite cutoff. This is shown in 
Fig. 1 for protons fixed at the ideal bcc lattice sites. The choice 
of the cutoff strongly depends on the average density: the lower $r_s$, the 
higher the cutoff. It is the number of plane waves which has to be kept
approximately constant in order to achieve similar levels of convergence.
In order to have the electronic density correctly described from less than
one third of the nearest-neighbor ionic distance onward, we decided to vary 
the cutoff between 230 Ry for $r_s=0.5$ and 60 Ry for $r_s=1$ and 1.2. It 
can be seen in Fig. 1 ($r_s=0.5$) that differences in the proton-electron
correlation function become noticeable at distances smaller than $0.5~a$,
while the nearest neighbor distance in the bcc lattice is $d_{nn}=1.76~a$. 
It should be pointed out that a cutoff of 60 Ry is sufficient to achieve 
convergence in the properties of the H$_2$ molecule, like the proton-proton 
distance (1.5 a.u. = 0.79~\AA) and the vibrational frequency (4160 cm$^{-1}$). 
The difference with respect to the experimental quantities (0.74~\AA~and 4400 
cm$^{-1}$) may be ascribed exclusively to the LDA. 

Isothermal ionic dynamics (particularly in the fluid phase) was achieved 
by using a Nos\'e-Hoover thermostat. The time step for integration of the
CP equations of motion was chosen typically between 0.25 and 1.5 atomic
units (depending on the temperature and density), i.e. roughly 10$^{-17}$ s,
compared to the ionic plasma period $\tau_P\approx 155~r_s^{3/2}$ a.u.
In some cases another Nos\'e thermostat was necessary to keep the fictitious 
kinetic energy of the electronic degrees of freedom at low values.
Most runs extended over 0.4 to 0.6 psec, which amounts to more
than 50 plasma oscillations. Before this, we allowed for thermal equilibration 
during an initial period of 0.2 psec in the solid phase, and 0.8 to 1.2
psec in the fluid phase. The densities $r_s=0.5$, $r_s=1$ and $r_s=1.2$ 
were explored in detail for temperatures in the range $100~K\leq T\leq 
10000~K$, which correspond to the strong coupling regime ($\Gamma\geq 30$), 
and cover both fluid and solid phases of the ionic component. In particular, 
$r_s\approx 1$ and $T\approx 7000~K$ are typical conditions inside Jovian
planets~\cite{hubbard}. The simulations at $r_s=1.2$ were carried out only
in the fluid phase because the stable solid has no longer the bcc symmetry, 
and thus it is not compatible with the choice of 54 H atoms in a simple 
cubic supercell.

\section{Validity of Linear Response Theory}

In the very high density limit ($r_s\to 0$) the electrons are barely 
polarized by the ionic charge distribution, due to their rapidly increasing
Fermi energy, and the hydrogen plasma practically reduces to two decoupled
components: a classical ionic ``one-component-plasma'' (OCP), and
a degenerate, rigid electron ``jellium''. Both systems have
been extensively studied by classical~\cite{baus} and quantum~\cite{cepald80}
computer simulations. For finite, but small $r_s$ ($r_s\ll 1$), the
coupling between the two components may be treated perturbatively within
linear screening theory~\cite{galam}. However, due to the strength of the 
unscreened Coulomb interaction between protons and electrons, such a
perturbative approach is expected to break down rapidly as $r_s$ increases.
In this section we investigate the high density limit and the validity
of the description of the hydrogen plasma in terms of a linear response
picture (LRT) for the electronic component.

To this end we compare the proton-electron pair correlation 
function taking into account the full LDA response:

\begin{equation}
g_{pe}(r)=-\frac{V}{N^2}\int d\bfk~\langle\rho_i(\bfk)~\rho_e(-\bfk)\rangle
~j_0(kr)
\end{equation}

\noindent and the one obtained by replacing the electronic charge
density by its linear response expression in terms of the ionic
charge density:

\begin{equation}
\rho_e(\bfk)=\chi_e(k)\left(-\frac{4\pi}{k^2}\right)\rho_i(\bfk)
\end{equation}

\noindent where $\chi_e(k)$ is the static electron density response
function. In eq. (2), $j_0(x)$ denotes the $l=0$ spherical Bessel function,
$\rho_i(\bfk)$ is a Fourier component of the microscopic ionic density,
and $\rho_e(-\bfk)=\rho_e^*(\bfk)$ is a Fourier component of the
electronic density corresponding to the instantaneous ionic configuration
$\{\bfrr_I\}$. The latter represents, within the adiabatic approximation, 
a ground-state expectation value.

  For the density response we adopt the random phase approximation (RPA) 
corrected by the long wavelength limit of the local field function
$G(k)$~\cite{baus}.

\begin{equation}
\frac{4\pi}{k^2}\chi_e(k)=\frac{k_{TF}^2~l(k/k_F)}
{k^2+k_{TF}^2~l(k/k_F)[1-G(k/k_F)]}
\label{pot}
\end{equation}

\noindent with $l(x)$ the Lindhard function and $k_{TF}$ the Thomas-Fermi
wavenumber. For the local field correction we use the following 
expression, which is consistent with the LDA and Slater's local exchange:

\begin{equation}
G(k/k_F)=\Big[\frac{1}{4}-\frac{\pi\lambda}{24}\left(r_s^3
\frac{d^2\epsilon_c(r_s)}{dr_s^2}-2r_s^2
\frac{d\epsilon_c(r_s)}{dr_s}\right)\Big]~\frac{k^2}{k_F^2}
\end{equation}

\noindent where the correlation energy density $\epsilon_c(r_s)$ is that 
proposed by Perdew and Zunger~\cite{pz}, and $\lambda=(4/9\pi)^{1/3}$.
At typical densities studied in this work the exchange contribution to
$G(k/k_F)$ turns out to be largely dominant over correlation, and both
of them represent a small correction relative to the RPA susceptibility.
In fact, this is reasonable because the RPA (or Lindhard approximation) 
is known to be good in the high density limit. 

The proton-electron distribution function reads accordingly:
 
\begin{eqnarray}
\lefteqn{g_{pe}^{LR}(r)= \frac{V}{N^2}\int d\bfk~
<\rho_i(\bfk)\rho_i(-\bfk)>~j_0(kr)}\nonumber \\ & \nonumber \\
&{\displaystyle\times\left(\frac{k_{TF}^2~l(k/k_F)}{k^2+k_{TF}^2~l(k/k_F)
[1-G(k/k_F)]}\right)}
\end{eqnarray}
 
In Fig. 2 we compare the proton-electron radial distribution 
functions in the ideal bcc structure as obtained from the LDA and from
LRT, for $r_s=0.5$ and $r_s=1$. As expected, $g_{pe}(r)$ is considerably 
more structured at the lower density. The LDA and LRT distribution 
functions are very close at $r_s=0.5$, but significant differences are 
clearly apparent at $r_s=1$, reaching a region well beyond the first ionic 
shell and signalling the break-down of the linear screening regime. When 
the ions are at finite temperature the differences are significantly 
enhanced, particularly at short distances. Also the location of the first 
minimum turns out to be shifted outwards by LRT in the fluid
phase (unlike in the solid phase), thus indicating that the LRT description
of the electronic charge distribution worsens for increasing temperature. 
This is shown in Fig. 3. The dashed curves have been obtained
by averaging the adiabatic electronic charge distribution around the protons
along the AIMD trajectory, assuming that the configurations generated in this 
way are also representative of the LRT. In fact, the magnitude of the 
differences observed in Fig. 3 suggests that the LRT trajectories will differ 
significantly from the LDA ones, implying that this averaged electronic
distribution might be meaningless within LRT. In other words, the 
fully-consistent LRT is likely to be worse than the rasults presented here.

Following a novel procedure~\cite{furio} we have fitted, for $r_s=1$, an 
effective proton-proton pair potential to our set of AIMD configurations 
generated at several temperatures spanning both solid and fluid phases. 
In Fig. 4 we compare this potential to the one obtained within LRT, i.e. 
by Fourier transforming $v_{\rm LRT}(k)=4\pi[1-\chi_e(k)]/k^2$, where 
$\chi_e(k)$ is given by expression~(\ref{pot}). The nearest neighbor 
distance in the bcc structure for $r_s=1$ is 1.76 a.u., and the position 
of the first peak in the proton-proton radial distribution function 
decreases down to values of the order of 1.5 a.u. in the fluid phase and 
upon heating (see below). This means that nearest pairs of protons will 
spend most of the time at distances of this order. Looking carefully at 
Fig. 4 it can be seen that, at those distances, the two potentials
differ by more than 10\%, the LDA one being steeper. On the other
side, the LRT pair potential appears to be more long-ranged than the
LDA one. This implies that the LDA pair potential has a characteristic 
screening length shorter than its LRT counterpart, i.e. screening is
more efficient than linear at $r_s=1$. 

The departure from the linear screening regime, also in the form of
many-body effects beyond the pair potential approach (e.g. three-body 
terms or embedding functions), becomes more pronounced as $r_s$ is 
increased, leading to ground state-atomic phases other than bcc, fcc or 
hcp (e.g. hexagonal and diamond) and, eventually, to recombination and 
to the different H$_2$ molecular phases. The reason for this
early departure from the LR regime has to be traced back to the unusual
strength of the bare Coulomb interaction, arising from the absence of 
core electrons. In fact, other alkali metals can be reasonably well described 
by LRT at much lower electronic densities ($r_s> 3$)~\cite{hafner}. 
Interestingly, the range of validity deduced here for the LRT is much 
wider than  expected in previous theoretical work based on
perturbative expansions~\cite{galam}, where $r_s=0.1$ was identified 
as the upper limit.

\section{Size effects and Fermi surface sampling}

In the solid phase, the combination of $\Gamma$-point sampling and a finite
system size is not expected to provide a very accurate description of
the electronic component, because all the periodicities beyond the size 
of the supercell are not properly included.
This is particularly important in metallic systems, where no point in the 
Brillouin zone can be taken as representative of the band structure of the
bulk solid. The reason is that occupied states (contributing to the electronic 
density) and empty states (which do not contribute) coexist in the same band,
corresponding to different k-points. A particular choice (e.g. the 
$\Gamma$-point and its refolded images) will sample the conduction
band in some specific points, but the character of the Fermi surface 
could be misinterpreted if empty states are taken as if they were occupied
and viceversa. For quasi-spherical Fermi surfaces this is unlikely to 
happen, but for transition metals or semimetals (like graphite) this effect 
is crucial, and a very fine sampling of the Brillouin zone is needed to  
obtain the right physics.

In the case of simple metals the effect of quantization 
of the electronic states in the simulation box is more important. The 
Brillouin zone of the unit cell is sampled with a finite number of points, 
which arise from the refolding of the $\Gamma$-point of the supercell. These 
points reflect the symmetries of the system in the sense that, if the 
$\Gamma$-point refolds onto some point $\bfk_0$, then all the points
in the star of $\bfk_0$ must arise from some other refolding of 
$\Gamma$. Otherwise, the symmetry is broken and spurious forces appear
that drive the system away from the symmetric ground state. Depending on 
the case at hand, the distortion can be rather large, especially if the 
symmetry of the supercell is very different from that of the unit cell. 
If the correct symmetry is used for the supercell then, since all the points
in the star are equivalent, the eigenvalues associated with them are
degenerate, thus giving rise to the formation of electronic shells.
If a better sampling of the Brillouin zone is performed, or the size
of the system is increased, new shells appear that eventually  
give rise to a continuous energy band. But for finite systems and restriction 
to the $\Gamma$-point, the electronic density of states consists of a set of 
discrete peaks (the shells). A problem arises with the highest occupied shell 
because in finite-size metallic systems it is partially occupied, unless 
a very fortuitous situation occurs. The immediate consequence is that it 
is not possible to fullfil the symmetry requirements by occupying all 
the points in the star with an {\it integer} number of electrons. Then, 
unless {\it fractional} occupation is introduced, the symmetry is broken 
even when the symmetry of the unit cell is mantained for the supercell.
This is analogous to the Jahn-Teller effect in molecular systems, where 
a lower-energy state can be obtained by reducing the symmetry and 
breaking the electronic degeneracy. However, in the present case
this effect is unphysical.

In a fluid phase these effects are normally expected to become less important
because of the breakdown of the discrete translational invariance, 
as reflected in the very existence of a finite Brillouin zone 
via Bloch's theorem. The unit cell becomes of infinite size and the 
Brillouin zone reduces to a single point, i.e. the $\Gamma$-point.
However, for practical reasons, computer simulations are bound 
to represent the infinite (fluid) system with a limited number of 
particles (typically of the order of some hundreds in ab initio calculations), 
while repeating periodically the supercell (via PBC). In fact, this 
introduces a spurious Brillouin zone, which is associated with periodicities 
that are absent in the infinite fluid. The properties of the fluid 
are assumed to be recovered from the finite sample in terms of PBC 
combined with statistical averages, which in the case of our AIMD 
simulations are computed as time averages.

This is a reasonable justification for a purely classical system and
also for liquid semiconductors but, for metallic systems, as soon as 
electronic states are introduced, the quantization of these states 
in the (unphysical) simulation box acquires a crucial role. 
The problem of partial occupation of the star persists in the fluid, 
in the sense that spurious forces appear that modify in a non-trivial 
way the structural and dynamical properties of the system.

A qualitative picture of the consequences of these observations can be 
obtained by comparing the proton-proton radial distribution function 
for systems of different sizes. A system of 54 H atoms (of valence charge
equal to 1) is really a fortuitous case of compatibility of a closed 
electronic shell (the whole star is fully occupied) with a bcc atomic 
arrangement in a simple cubic supercell. In the case of 128 H atoms only 
57 states are doubly occupied, and the next shell of 24 degenerate states 
has to accomodate 14 electrons (i.e. only 30\% of the shell). On the
contrary, 162 H atoms is not compatible with a bcc arrangement in a simple 
cubic supercell, but it closes the former electronic shell thus amounting
to 81 doubly occupied states. Therefore, in the fluid phase it should 
behave essentially as the 54-atom supercell but better converged in system 
size (or k-points). In Fig. 5(a) we show the curves corresponding to these 3 
different system sizes at $r_s=1$ and $T=1000~K$. In fact, up to the 
boundary of the 54-atom cell ($r=3~a.u.$) there is hardly any relevant
difference between the $g_{\rm pp}(r)$ corresponding to 54 and 162-atom
supercells. The same kind of picture holds also at higher temperatures,
meaning that, as regards static thermodynamic properties, 54
H atoms already give a very reasonable picture.

Very different is the situation with open shell systems like the 128 
H atoms one (short-dashed line in Fig. 5(a)). It is clear that this 
$g_{\rm pp}(r)$ has little to do with those of 54 and 162 atoms. The first 
peak is significantly lower and broader, and also its position is shifted 
downwards. The first valley is much shallower and, in practice, the pair 
distribution becomes structureless beyond it, while with 162 atoms the 
atomic shell structure is still visible at least up to the second valley. 
The 128-atom distribution resembles, in fact, the one that would have been 
obtained with closed shells at a higher temperature value. 

The correct way to account for open shell structures at finite temperature
is to doubly occupy the lowest-lying electronic states at every time step
of the AIMD simulation, while keeping empty the rest of that shell. 
The curves in Fig. 5(a) have been obtained by applying the standard procedure
of considering explicitly, in the description of the electronic component, 
a number of electronic states equal to one half of the number of electrons, 
i.e. including those strictly necessary. The justification for this approach 
is that typical temperatures are much lower than the Fermi temperature. 
Atomic motion leads to a degeneracy lifting of the order of a few percent of 
an eV, implying that the distance between electronic states is usually much 
larger than the width of the Fermi-Dirac distribution; this latter is of the
order of $k_BT$, i.e. about 0.1 eV at $T=1000~K$. This leads to a symmetry 
breaking in the sampling of the Fermi surface, which is expected to 
be recovered in terms of statistical averaging. However, following the 
approach above, in practical simulation times we have not noticed a 
convergence to the closed-shell picture. In Fig. 6 we show the Fermi-Dirac 
distribution function for three different temperatures, namely 1000 $K
$, 5000 $K$, and 10000 $K$, compared to the electronic density of states 
averaged along an AIMD run at $T=1000~K$, and to a snapshot of the 
instantaneous Kohn-Sham eigenvalues. It is clear that the occupation numbers 
fall from 2 to 0 in an energy scale narrower than the thermal splitting 
of the eigenvalues.

The drawback of this kind of approach is that for open shells the ordering
of the states within the highest (partially) occupied shell changes 
continuously during the MD evolution and, in particular, occupied states 
become empty and viceversa. The standard CP approach is not able to take into 
account this phenomenon, and this is one reason for the well-known failure
in metallic systems, reflected in the energy transfer between electronic
and ionic degrees of freedom (cooling the ions and heating the electronic 
orbitals). The double Nos\'e thermostat proposed by Bl\"ochl and 
Parrinello~\cite{bloechl} helps in fixing the temperature of each of
the components, but still does not take into account level crossing effects.
To our knowledge, there are three methods capable of solving this problem: 
one is to abandon the CP Lagrangian strategy in favor of a self-consistent 
minimization at each time step~\cite{payne}, a second one is to abandon 
the description of the electronic component in terms of single-particle 
orbitals to work directly with the electronic density which is, by definition,
constructed with the lowest occupied states~\cite{ali}, and finally a third 
one consisting of a rigorous Lagrangian formulation which incorporates the
occupation numbers (actually the conjugate variables, i.e. the Kohn-Sham
eigenvalues) as dynamical variables~\cite{hohlcar}. However, this latter
procedure exhibits the odd feature that the fictitious kinetic energy of
the electronic orbitals still increases at the expense of the ionic component, 
due to the appearance of low-energy excitations introduced by the dynamics of
the eigenvalues.

An approximate solution can still be found within the Lagrangian approach by 
evolving explicitly the whole open shell, but initially occupying -- with an 
integer number of electrons -- only the lowest states of the this shell. 
This will approximately take into account level crossing in terms of mixing 
of states within the shell, during the time evolution. The very same mechanism 
that leads to energy transfer between ions and electronic orbitals, i.e. the 
vanishing energy gap~\cite{pastore}, is responsible for mixing occupied and 
empty orbitals, thus allowing for initially empty states 
(if explicitly included) to become occupied and viceversa. Statistical 
averaging completes the task by generating a more uniform sampling of the 
Fermi surface. Fig. 5(b) shows how the results obtained with 128 H atoms 
can reproduce approximately those obtained with 54 and 162 H atoms. It has
to be pointed out that this procedure, although not rigorously justified,
has the nice feature that the fictitious kinetic energy of the electronic
orbitals is practically constant, behaving exactly as in systems with a 
gap, i.e. there is no energy exchange between ionic and electronic degrees
of freedom.

\section{Low-temperature phase: structure and melting}

In the OCP ($r_s\to 0$) limit the ionic bcc lattice is known to be the 
stable crystalline structure up to melting (which occurs at 
$\Gamma\approx 180$~\cite{baus}). We have studied the stability of the 
bcc structure at low temperatures and finite $r_s$ by performing canonical
MD simulations; initial conditions were constructed by giving the ions a 
small random displacement 
(about 3~\% of the nearest neighbor distance $d_{nn}$) from their alleged 
equilibrium positions (the bcc lattice sites). The bcc structure was found 
to be dynamically stable against such displacements at least up to $r_s=0.5$. 
At $r_s=1$ the bcc structure was found to be unstable for $T<100~K$, where 
a close-packed structure appears to be favored. This is consistent with a 
description in terms of LRT. In fact, at high densities the effective LRT
potential behaves essentially like a Yukawa potential 

\begin{equation}
v_{SC}(r)=\frac{1}{r}\exp(-r~k_{TF})
\label{yukawa}
\end{equation}

\noindent while Friedel oscillations are practically negligible. The phase 
diagram of a classical system of particles interacting via the above Yukawa 
potential~(\ref{yukawa}) has been extensively studied by computer simulations 
and lattice dynamics~\cite{robbins}. These calculations point to a bcc-fcc 
phase transition when the density-dependent screening wavenumber increases, 
i.e. when the effective interaction becomes of shorter range. At $T=0$ the 
bcc structure is found to be stable up to $r_s\approx 0.6$, beyond which 
the fcc phase becomes the stable structure. The $r_s$ at coexistence shifts 
to higher values at finite temperatures, such that the system goes through
a structural fcc-bcc phase transition as the temperature increases along 
an isochore. 

The situation here is reminiscent of the behavior of alkali metals. Na 
exhibits an hcp ground state, while Li goes from bcc to fcc and eventually 
to hcp at very low T. The heavier alkalis K, Rb and Cs undergo a 
structural transition to fcc upon cooling below $T\approx 5~K$. In
all these cases the entropic contribution of the bcc structure, arising
from the valley in the phonon dispersion along the (110) direction, 
wins over at finite temperature and stabilizes this phase. This is 
exactly what we observe in our simulations for H, where at $r_s=1$ the 
bcc structure appears to be stable for $T>100~K$.

Still, finite energy cutoff, finite system size, and coarse Brillouin 
zone sampling may have a large influence on the stability of the ground
state structure. The study of this part of the phase diagram deserves
special attention because also zero-point motion effects on the protons
have been shown to influence the stability of different structures at 
$T=0$~\cite{natoli2}. Disregarding the problem of zero-point energy (ZPE), 
and only as a check of the present calculations (which do not include ZPE), 
we have performed total energy full-potential Linear Muffin Tin Orbitals 
(FP-LMTO) calculations for solid, monoatomic H in the bcc, fcc and hcp 
structures, at $r_s=0.5$ and $r_s=1$. 
The energy differences turned out to be very small, but at $r_s=1$ the fcc 
and hcp structures are significantly lower in energy than the bcc. Moreover,
the energy differences are enhanced if a coarse sampling of the Brillouin 
zone is performed. These calculations also identify the hcp structure as
the lowest energy one, but the difference with respect to the fcc is within 
the accuracy of the calculations. The reason for this can be found in Fig. 
4. The differences between fcc and hcp structures begin only at the level 
of third nearest neighbors, a region where the effective potential shows 
negligibly small Friedel oscillations. 

Between $r_s=1$ and 1.2 a phase transition occurs that takes the system 
from hcp to a simple-hexagonal phase with a compressed $c/a$ ratio 
(squeezed-hexagonal). This is compatible with static total energy 
calculations by Barbee et al.~\cite{barbee}, and it is an additional 
confirmation of the breakdown of LRT, because pair potentials (like 
the LRT one) are not able to stabilize anisotropic structures like 
the simple-hexagonal. A more detailed study of the low-temperature 
atomic phases of Hydrogen is currently under way~\cite{ksct:unp}.

Next, the melting of the ionic crystal was investigated
by gradually increasing the temperature and monitoring the time-dependent
mean square displacement of the ions $<\mid\Delta\bfr(t)\mid^2>=
<\mid\bfr(t)-\bfr(0)\mid^2>$. In the crystalline phase,
$<\mid\Delta\bfr(t)\mid^2>$ goes over to $2<\mid\delta\bfr\mid^2>$ for
sufficiently long times, where $<\mid\delta\bfr\mid^2>=<\mid\bfr-\bfrr\mid^2>$
denotes the static mean square displacement ($\bfrr$ are the equilibrium 
positions of the ions). In the fluid phase diffusion sets in, so that 
$<\mid\Delta\bfr(t)\mid^2>=6Dt$ at long times, with $D$ the ionic diffusion 
constant. At $r_s=0.5$ diffusion
was found to set in at $\Gamma\approx 230$, which may be identified with
the limit of mechanical stability of the (overheated) metastable crystal.
The thermodynamic transition occurs at lower temperature (higher $\Gamma$);
its location may be estimated by assuming that the Lindemann ratio 
$L=(<\mid\delta\bfr\mid^2>)^{1/2}/d$ at melting is the same as for the
OCP, i.e. $L\approx 0.15$~\cite{hong}. This leads to 
$\Gamma_m(r_s=0.5)\approx 290$ compared to $\Gamma_m(r_s=0)\approx 180$,
indicating a strong influence of electron screening on the melting
transition. This confirms recent predictions based on free energy
comparisons, obtained by means of an approximate density functional 
theory~\cite{hong}. 

We have also studied the melting transition at $r_s=1$ using the same 
procedure; the melting temperature drops sharply from $T_m(r_s=0.5)\approx 
2200~K$ to $T_m(r_s=1)\approx 350~K$ ($\Gamma_m(r_s=1)\approx 930$). 
This indicates the possible proximity of a triple point bcc-hcp(fcc)-liquid,
analogous to that found for Yukawa potentials. However, the nonlinearity of 
the screening at these values of $r_s$ is likely to bring the triple point 
from $r_s=3$~\cite{robbins} to $r_s\approx 1.1$. Moreover, the hcp structure 
goes over to a simple hexagonal one at $r_s$ somewhere between 1 and 1.2, 
and this has to be a consequence of the appearance of anisotropic forces, 
beyond the level of pair-wise additivity. The very low value of the melting
temperature might also be related to the appearance of these forces.
Interestingly, at $r_s=1$ the Fermi temperature of the ionic component is 
$T_F^p=326~K$, a value close to the melting point ($350~K$). Therefore, 
the influence of quantum effects for the protons on the melting transition 
cannot be ignored, and will probably also play an important role in the 
above structural phase transition. In particular, they might destabilize 
the hexagonal phase in favor of some more isotropic configuration which 
has a higher energy within a framework of classical protons.

\section{The fluid phase: an atomic-like plasma}

Turning to the fluid phase, a quantitative measure of ion-electron
correlations is provided by the sphericalized average of the 
ion-electron pair correlation function $g_{pe}(r)$, as computed from eq. (2).
The effect of temperature on $g_{pe}(r)$ is illustrated in Fig 7, 
for $r_s=0.5$ and $r_s=1$. The distribution function is seen to be
remarkably insensitive to $T$ over the whole range of temperatures, 
covering the solid and fluid phases, as already noticed at lower 
$\Gamma$ (higher $T$) by Dharma-Wardana and Perrot in the 
framework of an approximate static DFT-HNC
calculation~\cite{dharma}. The observed weak temperature dependence implies
that the main effect of ionic thermal motion (electrons are always at $T=0$), 
is to enhance the  localization of the electronic charge close to the protons.

This behavior is to be contrasted with the predictions of higher-level 
theories that go beyond the Born-Oppenheimer approximation by including 
excited electronic states. Both, fully-quantum PIMC~\cite{pierleon} and Mermin 
functional~\cite{alavi} simulations imply that ion-electron correlations become
weaker as temperature increases. This is to be intuitively expected, but for 
temperatures much higher than the ones studied here. Excited electronic 
states are much more insensitive to ionic polarization effects, because they
correspond to larger kinetic energies. In this way, for temperatures larger
than a threshold value that can be estimated around $\theta=0.1$, i.e. 
$T\approx 60000~K$ for $r_s=1$, the above localization effect due to ionic
disorder starts to be compensated by the effect of electronic excitations, so 
that eventually the opposite trend will take over.
 
The ion-electron pair correlation functions for {\it all}
temperatures are seen to intersect at a well-defined (reduced) distance 
from the proton site, irrespective of thermal ionic disorder and almost 
independently of density. We locate this value at $r^*\approx 1.3~a$,
and notice that the ratio $r^*/d_{nn}\approx 0.73$ is related to the
ratio of the location of the nodes corresponding to the first two spherical 
Bessel functions ($j_0(x)$ and $j_1(x)$). In fact, the electronic problem
can be modelled, in a very crude approximation, as that of a particle in a 
spherical well; the corresponding radial solutions are precisely the spherical 
Bessel functions. Temperature effects can be mimicked by increasing the
relative population of excited states with respect to $j_0(x)$. However,
the location of the node of $j_1(x)$ (the leading excitation) 
relative to $j_0(x)$ (the ground state) does not depend on temperature.
Since the location of the nodes is defined in units of the radius of the
well, and this is identified with $d_{nn}$, which is proportional to $r_s$, 
the crossing should not depend significantly on density. The first maximum 
of $g_{pe}(r)$ is clearly associated with the location of the first 
coordination shell (the first maximum in $g_{pp}(r)$ -- see below), 
which is quite natural since the electronic density peaks at the 
proton sites. It is interesting to notice that differences in the electronic
screening properties between $r_s=1$ and $r_s=1.2$ are significative
only in the vicinity of the protons, up to $r\approx 0.5~a$.

The proton-proton pair correlation function for $r_s=1$ is illustrated in 
Fig. 8(a), as a function of temperature. It can be observed that the first 
peak remains clamped at the nearest neighbor distance ($d_{nn}=1.76$) for 
temperatures below the mechanical stability limit -- $T_s(r_s=1)\approx 500 
K$ --, while in the fluid phase it shifts continuously to shorter distances. 
The same plot shows that the location of the first minimum is 
quite insensitive to temperature. This, together with the fact that 
equivalent results are found at $r_s=0.5$, defines quite univocally a first 
coordination shell of radius $r_{fcs}\approx 2.4~a$ (with $a$ the ion-sphere
radius). The integrated number of particles is shown in Fig. 8(b) as 
a function of temperature. The main result is that the first coordination 
shell contains 14 atoms on average, implying that the short-range structure 
of the liquid is quite reminiscent of that of the solid, since the first 
coordination shell of the bcc structure (containing first and 
second nearest neighbors) also contains 14 atoms. Simulations performed
with 162 H atoms show that this is a genuine feature and not an 
artifact of the small system size. A second coordination shell is also
well-defined in the fluid phase provided that the temperature is low
enough, i.e. $T<1500~K$ ($\Gamma>200$), as can be observed in Fig. 9. 
However, the fluid becomes structureless beyond the first coordination 
shell at temperatures of the order of 5000~K (at $r_s=1$). Summarizing, 
the fluid phase of the H plasma at moderately high temperatures and very 
high densities (typical of the inner H shell of Jovian planets) behaves 
like an atomic liquid with a well-defined first coordination shell.

The influence of electron screening is also apparent when comparing the
ion-ion and charge-charge static structure factors $S_{ii}(k)$ and 
$S_{ZZ}(k)$. While at $r_s=0.5$, the two are nearly indistinguishable, 
the amplitudes of their main peaks differ significantly for $r_s=1$ (by
roughly 9~\%), and $r_s=1.2$ (12\%). Due to the discrete sampling of the
electronic density in reciprocal space, the curves are noisy and will not
be reproduced here. The strong polarization of the 
electronic component leads to a damping of the local charge fluctuations, 
and hence to a reduction of $S_{ZZ}(k)$. Again, the presence of a 
well-defined first peak and valley in $S(k)$ is an indication that the 
fluid is well structured in this region of the parameter space.

\section{Diffusion coefficients and Vibrational properties}

Our AIMD simulations give direct access to the ionic dynamics by analysing 
the time evolution of atomic coordinates and velocities. Diffusion 
coefficients have been calculated using the asymptotic relation 
$<\mid\Delta\bfr(t)\mid^2>=6Dt$; i.e. by measuring the slope of the mean
square displacement of the atoms as a function of time. Our results are 
displayed in table I for $r_s=0.5$,$r_s=1$ and $r_s=1.2$, in reduced plasma 
units $D^*=D/a^2\omega_{\rm pl}$, where $\omega_{\rm pl}(r_s)=(3/M_I)^{1/2}~
r_s^{-3/2}$ is the bare ionic plasma frequency. The results are shown in 
a log-log plot in Fig. 10. The relationship between $D^*$ and 
$\Gamma=1/(r_sT)$ follows quite accurately a power-law of the type 
$D^*=D_o\Gamma^\alpha$. We have fitted our data to such an expression, 
obtaining for $r_s=1$ the following values: $D_o=(10.4\pm 1.4)$ and 
$\alpha=(-1.38\pm 0.07)$, and for $r_s=0.5$ the values: $D_o=(4.0\pm 1.5)$ 
and $\alpha=(-1.37\pm 0.09)$. 

It is interesting to note that the diffusion coefficient follows the
same relationship as in the OCP model. The OCP values for the parameters,
fitted to classical MD simulations~\cite{pollock} are $D_o=2.95$ and
$\alpha=-1.33$. The value of the exponent seems to be unaffected by 
electronic screening, at least within the accuracy of our calculations. 
The prefactor, however, is clearly enhanced from its OCP value, and this 
can be readily understood in terms of the response of the electronic 
component to the motion of the protons.  A rigid uniform electronic 
background (as in the OCP) does not have any influence on the dynamics of the 
protons. A polarizable background weakens the proton-proton interaction 
thus increasing the mobility of the ions. In fact, 
the results obtained at $r_s=0.5$ are quite close to the OCP values,
represented by the dashed line in Fig. 10. The difference becomes much 
larger at $r_s=1$, where $D^*$ differs from its OCP values by a factor of 3; 
this contrasts with the Thomas-Fermi MD results of  Z\'erah et 
al.~\cite{zerah}, who observed a much milder effect (a factor 1.4 at 
$\Gamma=50$ and $r_s=1$). Therefore, the diffusion coefficient appears to 
be very sensitive to the treatment of electron screening. The case of 
$r_s=1.2$ is slightly different, because the exponent appears to be larger 
than the OCP value. However, it can be seen that the error bars are also 
compatible with an OCP-like power-law, represented by a line parallel to 
that of the other densities.

The diffusion coefficient is also related to the ion velocity 
autocorrelation function (VACF), which, in the OCP limit, exhibits a 
striking oscillatory behavior due to a strong coupling of the 
single-particle motion to the collective ionic plasma 
oscillations~\cite{pollock}. Such oscillations were recently shown 
to persist at finite $r_s$, by MD simulations using the approximate 
Thomas-Fermi kinetic energy functional instead of the Kohn-Sham 
version~\cite{zerah}. The present ab initio calculations qualitatively 
confirm this behavior. In Fig. 11 we present the VACF for a typical
simulation in the solid phase ($T=300~K$) and then for three different
temperatures in the fluid phase. The latter were computed in
the supercell containing 162 atoms. Finite size effects are not very
significant as regards the general features of the VACF. However,
a small frequency shift is observed, and decorrelation happens faster
in the larger sample. It is interesting to note that the fastest 
oscillation, i.e. the one associated with the ion plasma oscillations, is
essentially temperature-independent. 

The power spectra of the ionic VACF are plotted in Fig. 12 for $r_s=1$, 
at several temperatures. The spectra exhibit a high-frequency 
peak (or shoulder at the highest temperature) at a frequency 
which amounts to 55~\% of the bare ion plasma frequency ($\omega_{\rm pl}$).
The power spectra for $r_s=0.5$ are similar, with the difference that the
high-frequency peak occurs now at a value which is 70~\% of the bare plasma
frequency~\cite{kh:prl}. As expected, electron screening shifts the 
vibrational spectra to lower frequencies. Temperature, however, does not
affect the position of the high-frequency peak, which remains the only
well-defined feature at high temperatures, while the rest of the spectrum 
merges into a structureless continuum. These spectra were obtained from 
54 H atom simulations. The 162 H atom sample yields a frequency 5~\% lower
than the 54 H atom one.

\section{Collective modes: a signature of the metal-insulator transition}

The oscillations in the VACF point to a long-lived longitudinal 
collective mode, related to the ionic plasmon mode of the 
OCP~\cite{pollock}. We have computed the charge autocorrelation function:

\begin{equation}
F_{ZZ}(\bfk,t)=\frac{1}{N}<\rho_Z(\bfk,t)~\rho_Z(-\bfk,0)>
\end{equation}

\noindent where the Fourier components of the microscopic charge density 
are:

\begin{equation}
\rho_Z(\bfk,t)=\rho_i(\bfk,t)-\rho_e(\bfk,t)
\end{equation}

\noindent with $\rho_i$ and $\rho_e$ the AIMD-generated time-dependent 
densities. In keeping with the Born-Oppenheimer approximation, 
$\rho_e(\bfk,t)$ is a Fourier component of the expectation value of the 
electronic density for the instantaneous ion configuration. In practice, 
we computed the average of $F_{ZZ}(\bfk,t)$ over the shell of 
equal-modulus $\bfk$-vectors

\begin{equation}
F_{ZZ}(k,t)=\sum_{\vert\bfk\vert=k}~F_{ZZ}(\bfk,t)
\end{equation}

\noindent From this, we computed the dynamical structure 
factor $S_{ZZ}(k,\omega)$ by Fourier transforming $F_{ZZ}(k,t)$. In the 
$r_s=0$ (OCP) limit, where the electrons form a uniform (non polarizable) 
background, $S_{ZZ}(k,\omega)$, which there reduces to the dynamical 
structure factor of the bare ions, is simply the k-dependent spectrum of 
the ionic plasma oscillations~\cite{pollock}. In the long wavelength 
limit the mode is undamped, and its characteristic frequency is the ion 
plasma frequency $\omega_{\rm pl}$. Adiabatic electron polarization 
transforms this ionic plasmon (or optic) mode into an acoustic mode for 
any finite value of $r_s$~\cite{postogna,barrat}. This mode is to be 
identified with the familiar low-frequency ion-acoustic mode.
Only if the system were treated as a fully dynamical ion-electron plasma, 
would the high-frequency plasma oscillation mode appear, which is related 
to the fast electronic motions. This mode is obviously not accessible by 
adiabatic MD simulations.
The conjectured scenario is confirmed by the results of our AIMD 
simulations. The dynamical structure factor $S_{ZZ}(k,\omega)$ was 
computed for the smallest wavenumber k compatible with the PBC 
($ka=1.031$ for the 54 atom system) and for selected larger wavenumbers 
($ka<3$). The resulting $S_{ZZ}(k,\omega)$ for $r_s=0.5$, $r_s=1$ and 
$r_s=1.2$ are shown in Fig. 13, for the smallest available wavenumber, 
and at a temperature just above melting. The sharp peaks are characteristic 
of the long-lived (weakly damped) mode anticipated above. The peaks shift 
to lower frequencies as $r_s$ increases due to enhanced electron 
screening, and in accord with the behavior of the plasmon-like peak 
in the spectrum of the VACF (the single-particle excitation coupled to 
the collective plasmon mode). 
The k-dependence of the spectrum $S_{ZZ}(k,\omega)$ is illustrated in Fig. 
14, for $r_s=1$; the resulting dispersion curve is shown in fig. 15. A 
striking feature is the nearly constant width of the resonance peaks for 
$ka<1.5$, pointing to a nearly k-independent damping mechanism. The damping 
increases dramatically at larger wavenumbers, while the dispersion curve 
bends over; the behaviour is reminiscent of that observed for a classical 
fluid of atoms interacting via an effective Yukawa potential~\cite{barrat}. 
The bending over may be regarded as a remnant of the negative dispersion of 
the plasmon mode observed in the strongly-coupled OCP~\cite{pollock}. From 
the initial slope of the dispersion relation, we estimate a sound velocity 
of $c_s\approx 70$ km/s at $r_s=1$, which is is consistent with the 
extrapolation of very recent results by Alavi et al. to the ultra-high 
density regime~\cite{aliala}. Sound velocities are relevant to the 
determination of global free oscillations of Jovian planets, which have
been recently measured for Jupiter~\cite{mosser}.

The remarkable feature is that this collective mode is much sharper 
than the usual sound mode observed in metals at comparable wavenumbers. 
Moreover, the peaks do not shift significantly with temperature, 
although they broaden. However, particularly
for low values of $k$, the signature of the collective mode can be
detectable up to quite high temperatures (well above 3000 $K$). This
is seen in Fig. 16, where we plot the dynamical structure factor for
$ka=1.238$ (i.e. just before the bending-down point) in the 162-atom 
sample at $r_s=1$ and for three different 
temperatures. It is interesting to note that this is precisely the 
temperature range where a transition is expected to occur to the 
molecular (H$_2$) fluid phase at lower densities~\cite{weir}. 
The observed collective behavior may be regarded as characteristic 
of the metallic phase of hydrogen, and is expected to change 
dramatically at the transition towards the molecular phase, which
begins to show up at low temperatures at $r_s\approx 1.3$. Thus, an 
analysis of 
$S_{ZZ}(k,\omega)$ may provide an efficient diagnostic to locate the 
plasma phase transition at finite temperature.

\section{Conclusions}

The main conclusions to be drawn from the present AIMD simulations of the 
hydrogen plasma in the high-density ($r_s\leq 1.2$) regime may be 
summarized as follows:

a) Due to the significant spacing between the quantized electronic states 
in the vicinity of the Fermi surface, the N-dependence of the statistical 
averages must be treated with great care, in order to extract meaningful 
results.

b) A linear-response treatment of the ion-electron correlations yields 
reasonable results at $r_s=0.5$, but becomes rapidly unreliable at lower 
densities.

c) The bcc structure, which is the stable low temperature solid phase at 
least up to $r_s=0.5$, becomes unstable at lower densities, where hcp 
and simple-hexagonal phases appear. More work is needed to determine 
the full low-temperature phase diagram, also including zero-point-motion
effects.

d) The melting temperature drops sharply with decreasing density, due to 
the enhanced efficiency of electron screening of the effective interaction 
between ions. New interesting physics is likely to arise in the region
of $r_s\approx 1.1$ and $T\approx 100~-~200~K$, where the existence of 
a bcc-hcp(fcc)-liquid triple point is argued, in a region where quantum
effects in the protons are non-negligible.

e) The fluid metallic phase behaves very much like a simple atomic liquid 
from a structural point of view, but the longitudinal collective dynamics 
of the ions retain a strong plasma-like character at intermediate 
wavenumbers. This reflects itself in unusually sharp peaks in the 
charge-fluctuation spectrum, which are gradually shifted to lower 
frequencies with decreasing density, as a result of electron screening. 
The damping, however, appears to be surprisingly insensitive to density, 
but is significantly enhanced by temperature. The acoustic character of 
the longitudinal mode is recovered at sufficiently small wavenumbers 
($ka<1$), in qualitative agreement with a simple linear screening picture. 
A strong damping of the mode at intermediate wavenumbers should be a 
clear-cut signature of the plasma-to-molecular phase transition, which 
is expected to start at $T=0$ around $r_s=1.3$, and to move to finite
temperatures of the order of a few thousand $K$ at lower densities 
($r_s>1.3$)~\cite{weir}.

f) The single-particle motion of the ions couples to the longitudinal 
collective mode, and reflects itself in a striking oscillatory behaviour 
of the velocity autocorrelation function, which is reminiscent of the 
behavior of the OCP, despite the action of strong electron screening. 
The resulting ionic self-diffusion constant is strongly enhanced at 
lower densities, for identical values of the plasma coupling constant 
$\Gamma$, but follows a power law similar to that observed in the OCP.

The present AIMD simulations will be extended to lower densities, in order 
to characterize the plasma-to-molecular phase transition, starting from 
the high density, metallic side.

\acknowledgments
One of us (JK) would like to thank Furio Ercolessi for facilitating 
his code to fit the two-body potential, and Ruben Weht for helping with
the FP-LMTO calculations. We acknowledge helpful discussions with Ali 
Alavi, Detlef Hohl, Hong Xu, Gilles Zerah, Stephane Bernard, Carlo 
Pierleoni, Pietro Ballone, Erio Tosatti, Giorgio Pastore and Sandro 
Scandolo.
%--------------------------------------------------------------------------

%--------------------------------------------------------------------------
%section*{Figure Captions}
%
\begin{figure}
\caption{Electron distribution around the proton site for H 
in the perfect bcc structure at $r_s=0.5$, as a function of the plane
wave energy cutoff. Dotted, dash-dotted and solid curves correspond
to cutoffs of 60, 230 and 420 Ry, respectively.}
\end{figure}
\begin{figure}
\caption{Electron distribution around the proton site for H 
in the perfect bcc structure at $r_s=0.5$ (lower curves) and $r_s=1$ 
(upper curves), according to LDA (solid lines) and within linear 
response (dashed lines). Distances expressed in units of the 
ion-sphere radius $a=r_s~a_B$.}
\end{figure}
\begin{figure}
\caption{Electron distribution around the proton site for H at 
T=2000 K according to LDA (solid line) and LRT (dashed line), for
$r_s=1$ (lower curves) and $r_s=1.2$ (upper curves).}
\end{figure}
\begin{figure}
\caption{Effective pair potentials for the proton-proton 
interaction at $r_s=1$: $v_{\rm LDA}$ fitted to AIMD trajectories (solid 
line), and $v_{\rm LRT}$ obtained within the RPA with local field corrections 
(dashed line).}
\end{figure}
\begin{figure}
\caption{Proton-proton pair distribution function at 
$T=1000~K$ and $r_s=1$ for different system sizes: (a) 54 atoms (solid line), 
128 atoms (dashed line), and 162 atoms (dotted line); (b) 128 atoms 
considering explicitly the whole open shell (81 states), where the lowest
64 are {\it initially} occupied (solid line), 128 atoms considering 
explicitly only the 64 lowest occupied states (dashed line), and 162 
atoms (dot-dashed line).}
\end{figure}
\begin{figure}
\caption{Fermi-Dirac distribution function for $T=1000~K$ (solid line),
$T=5000~K$ (long-dashed line), and $T=10000~K$ (dot-dashed line). The 
electronic density of states associated with the last two occupied shells
is also shown (short-dashed line), as well as the single shot of the
Kohn-Sham eigenvalues during an AIMD simulation at $T=1000~K$ (vertical
bars).The EDOS is broadened with a Gaussian window 1 eV wide.} 
\end{figure}
\begin{figure}
\caption{Proton-electron pair correlation function at finite
temperature (solid and fluid phases) at $r_s=0.5$ and
$T=1000, 2500, 5000~{\rm and}~10000 K$ (lower curves), and at $r_s=1$ and
$T=200, 500, 1900~{\rm and}~7300 K$ (upper curves). The lower curve of each 
set corresponds to the lower temperature.}
\end{figure}
\begin{figure}
\caption{(a) Proton-proton pair distribution function at 
$r_s=1$ and $T=300$, 500, 1000, 1500, 2000, 2500 and 7300 $K$. (b) 
Integral of the above distribution function for the same temperatures. The
function $N(r)$ indicates how many protons are contained in a sphere of 
radius $r$, on average.}
\end{figure}
\begin{figure}
\caption{Proton-proton pair distribution function at 
$r_s=1$ for a 162 H atom sample, as a function of temperature: $T=1000~K$ 
(solid line), $T=2000~K$ (dashed line) and $T=3000~K$ (dot-dashed line).}
\end{figure}
\begin{figure}
\caption{Reduced diffusion coefficient at $r_s=0.5$ (circles), $r_s=1$
(triangles), and $r_s=1.2$ (diamonds) in a log-log plot. Solid lines are
the results of a linear regression fit on the logarithms. The dashed
line is the OCP result.}
\end{figure}
\begin{figure}
\caption{Velocity autocorrelation function at $r_s=1$
for $T=300~K$ (dotted line), $T=1000~K$ (dashed line), $T=2000~K$ 
(dot-dashed line), and $T=3000~K$ (solid line). The abscisa (time) is
expressed in units of the plasma frequency $\omega_{\rm pl}$.}
\end{figure}
\begin{figure}
\caption{Power spectra of the velocity autocorrelation 
function at $r_s=1$, obtained by FT the velocity autocorrelation function.
The curve at $T=300~K$ (solid line) corresponds to the solid phase, while
the remaining curves correspond to $T=800~K$ (dotted), $1000~K$ 
(short-dashed), $2000~K$ (long-dashed), $2500~K$ (dot-dashed) and $7300~K$ 
(solid). These are all in the fluid phase. Frequencies are in units of the 
ionic plasma frequency $\omega_{\rm pl}$. Curves are normalized to integrate 
to the number of ionic degrees of freedom.}
\end{figure}
\begin{figure}
\caption{Charge-charge dynamical structure factor at the 
smallest available wave vector, $ka=1.031$, for three values of $r_s$.}
\end{figure}
\begin{figure}
\caption{Charge-charge dynamical structure factor at $r_s=1$ for different 
values of momentum: $ka=0.715$ (solid line), 1.011 (dotted), 1.238 
(short-dashed), 1.599 (long-dashed), 2.022 (dot-dashed), 2.476 (diamonds),
and 2.948 (crosses), from a simulation with 162 H atoms. The curves 
have been obtained by Fourier transforming the autocorrelation functions (8), 
filtering with an appropriate decaying exponential those corresponding to 
the largest wavevectors in order to reduce noise. All curves are normalized 
by the static structure factor $S_{ZZ}(k)=\int_0^\infty S_{ZZ}(k,\omega)~
d\omega$, such that the area under every curve is 1.}
\end{figure}
\begin{figure}
\caption{Dispersion relation for the collective ion acoustic mode in 
$S_{ZZ}(k,\omega)$, at $r_s=1$. Filled circles are points corresponding 
to the $k$-vectors allowed by the PBC, in a simulation box containing 
162 H atoms. The solid line is a guide to the eye, arising from a 
4$^{\rm th}$-degree polynomial fit to the data.}
\end{figure}
\begin{figure}
\caption{Dynamical structure factor at $r_s=1$ for $ka=1.238$ at $T=1000~K$,
(dot-dashed line), $T=2000~K$ (solid line), and $T=3000~K$ (dashed line).}
\end{figure}

\begin{table}
\caption{Diffusion coefficient as a function of the plasma coupling 
parameter $\Gamma$. The error in the determination of the diffusion 
coefficients is $\Delta D^*=0.001$.}

\begin{tabular}{c|c|c|c}
 $r_s$ & $\Gamma$ &  $D^*$  &  T (K) \\ \hline
 0.5 &   65   &   .0122  & 10000 \\ 
 0.5 &   90   &   .0090  &  7500 \\ 
 0.5 &  130   &   .0052  &  5000 \\ 
 0.5 &  210   &   .0025  &  3100 \\ \hline
 1.0 &  108   &   .0163  &  3000 \\
 1.0 &  130   &   .0119  &  2500 \\
 1.0 &  163   &   .0096  &  2000 \\
 1.0 &  217   &   .0066  &  1500 \\
 1.0 &  326   &   .0034  &  1000 \\ \hline
 1.2 &  136   &   .0173  &  2000 \\
 1.2 &  181   &   .0114  &  1500 \\
 1.2 &  217   &   .0072  &  1000 \\
 1.2 &  272   &   .0056  &   800 \\ 
\end{tabular}
\end{table}

%--------------------------------------------------------------------------

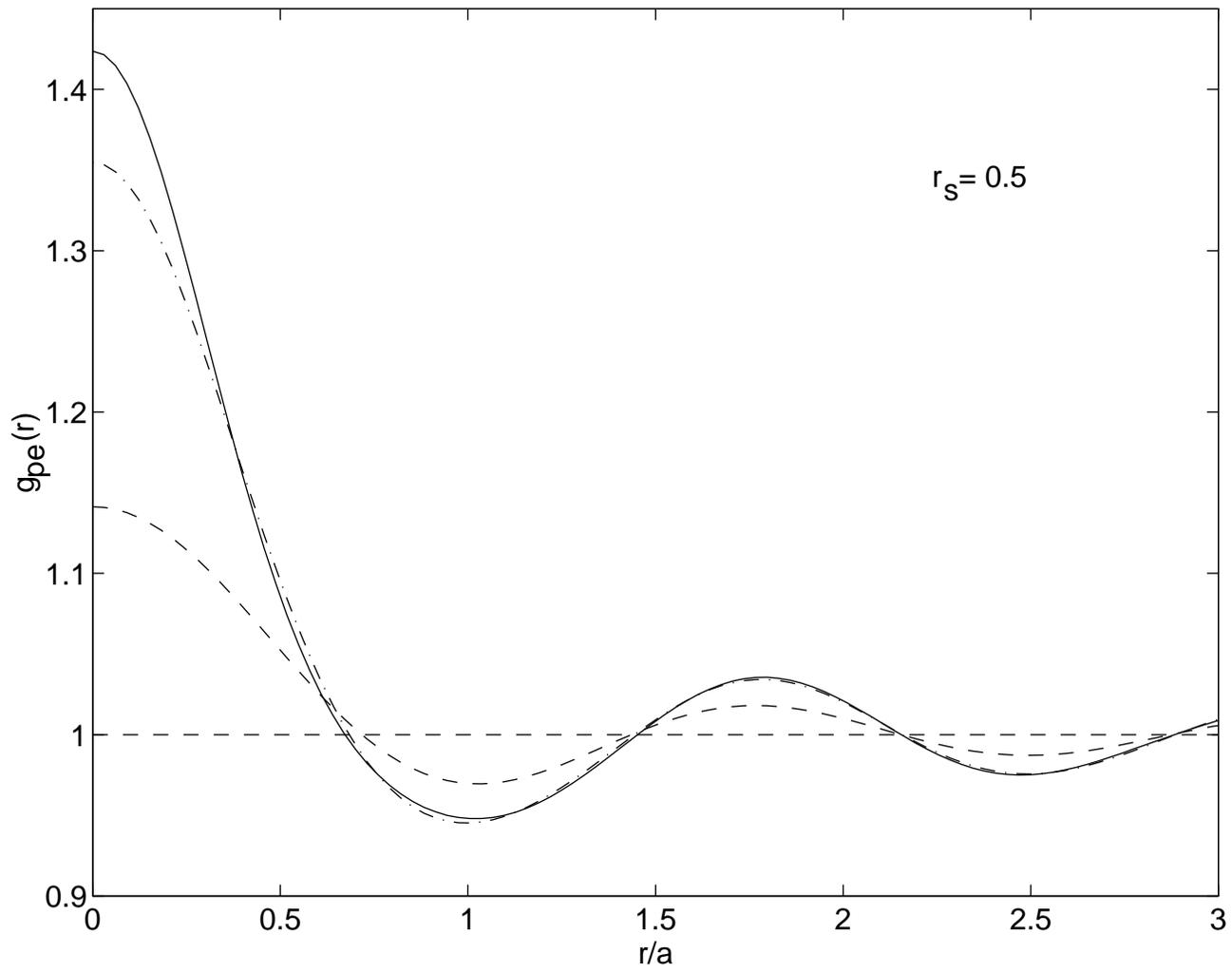
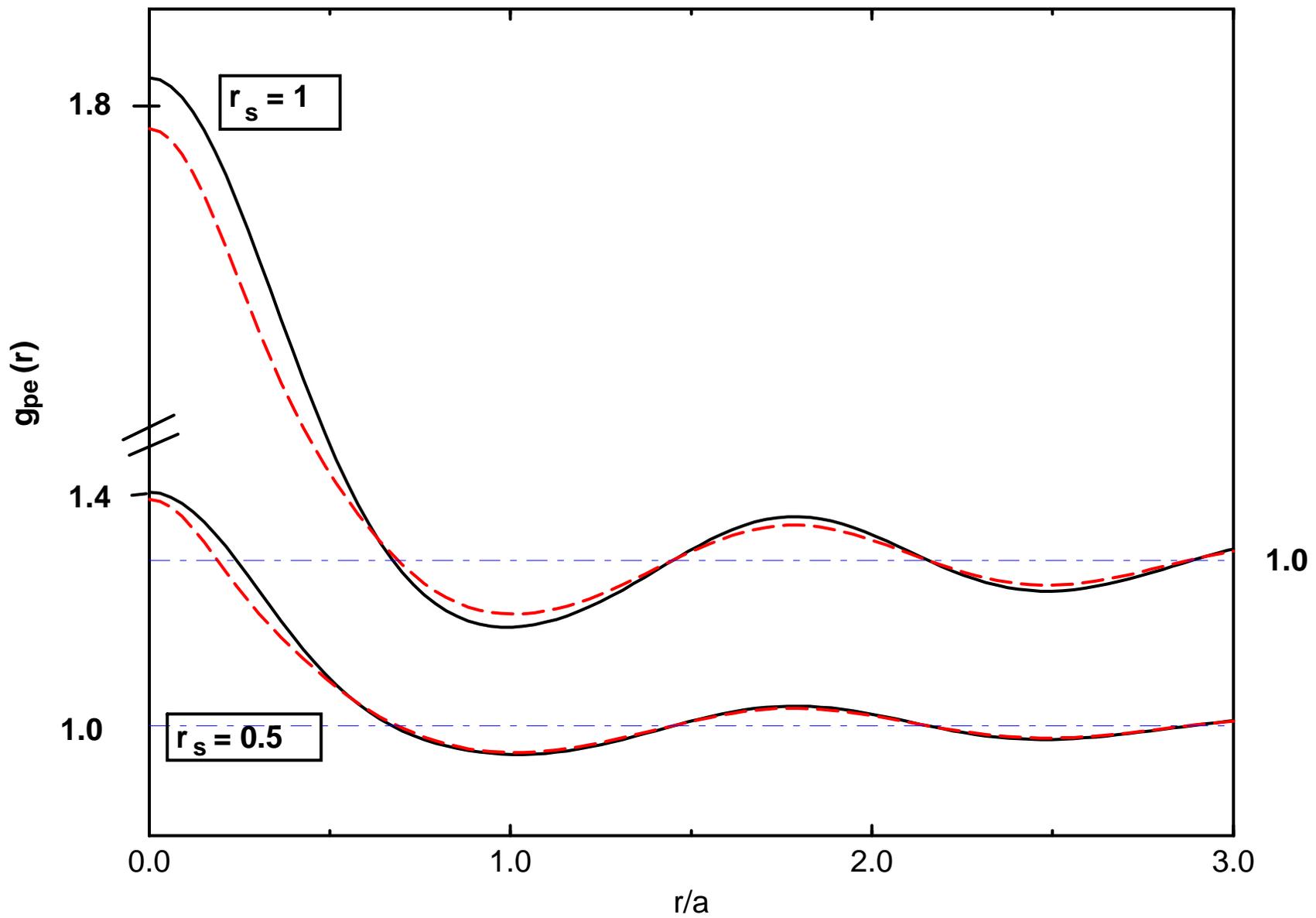
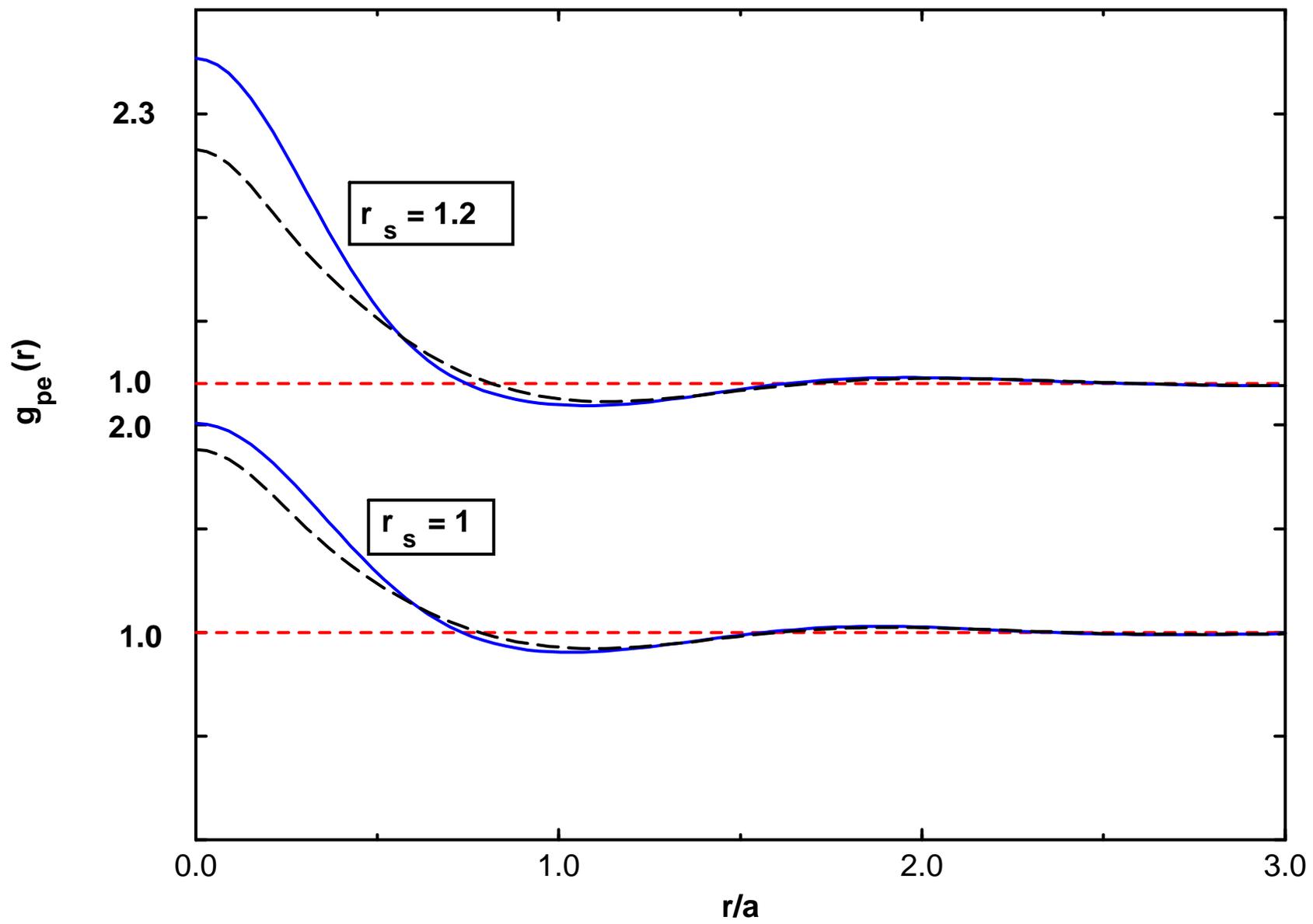
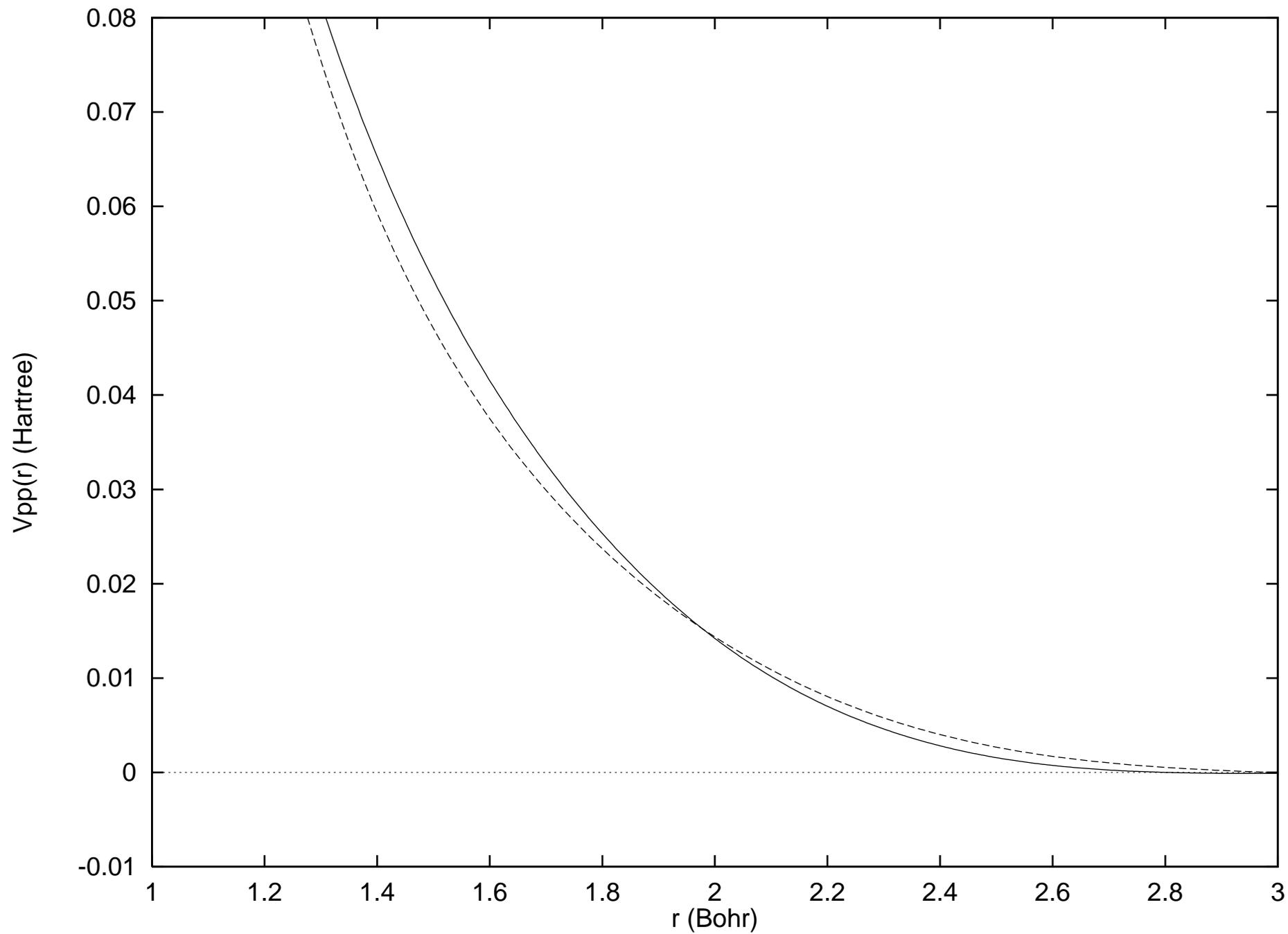
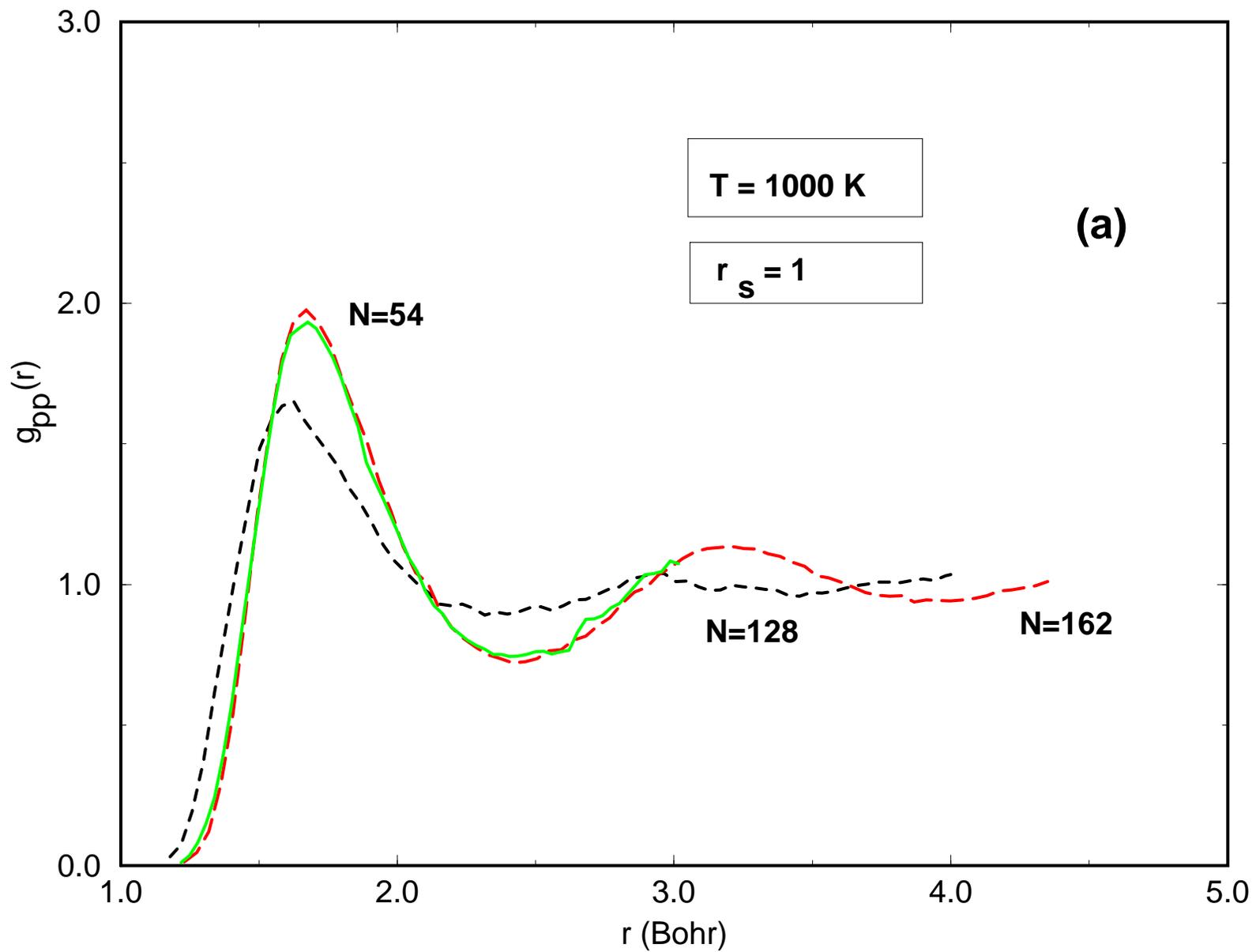
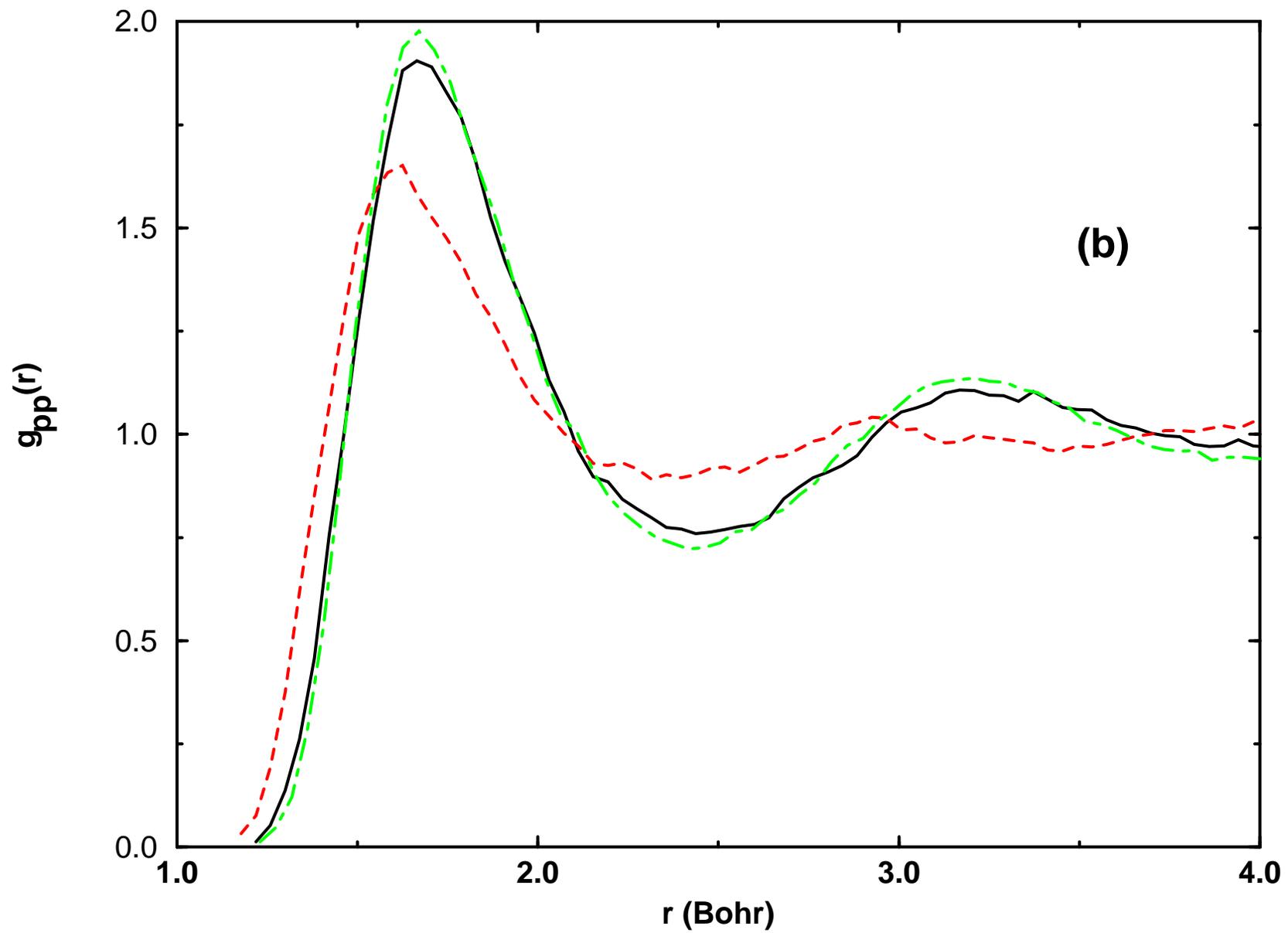
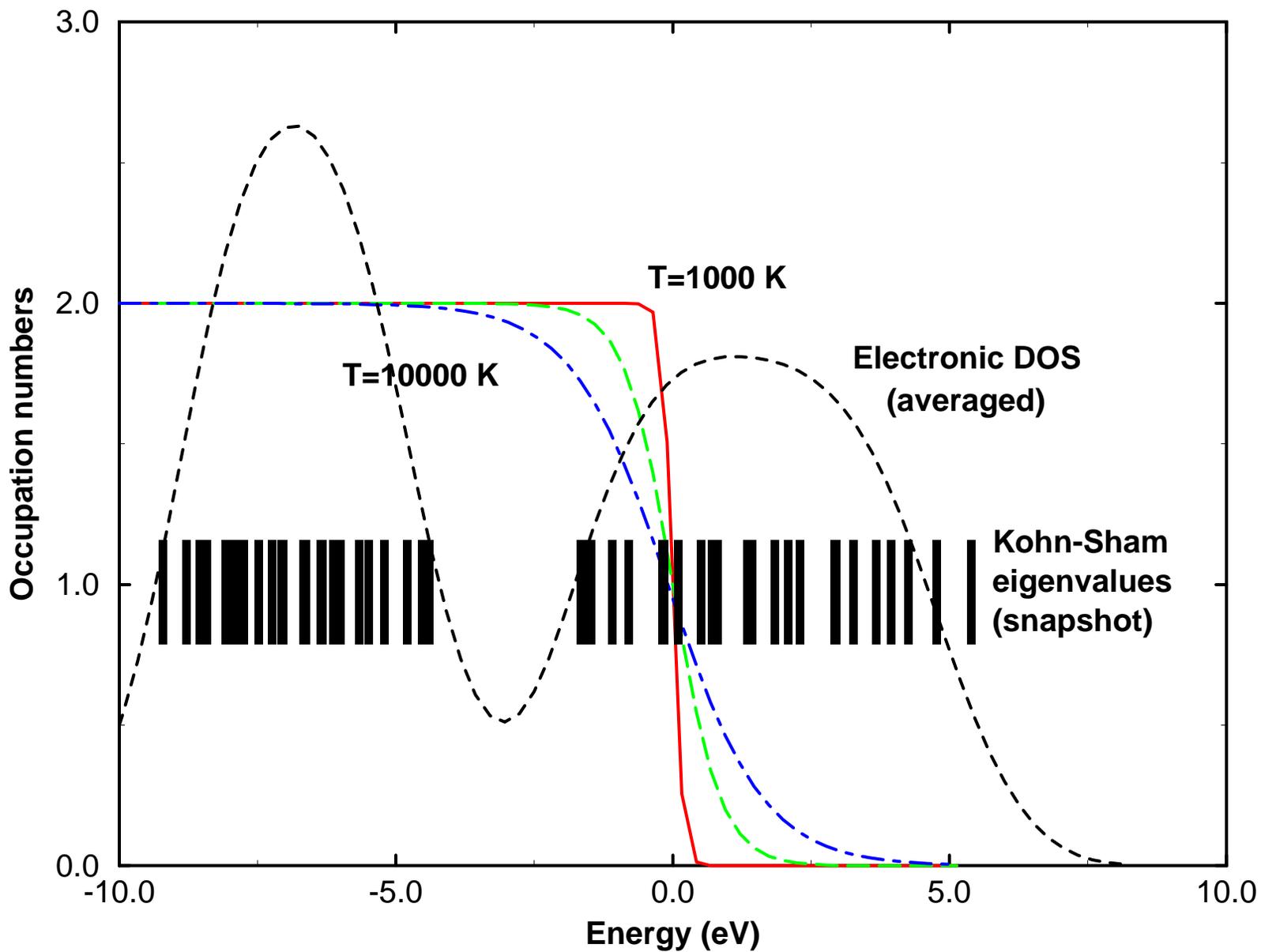
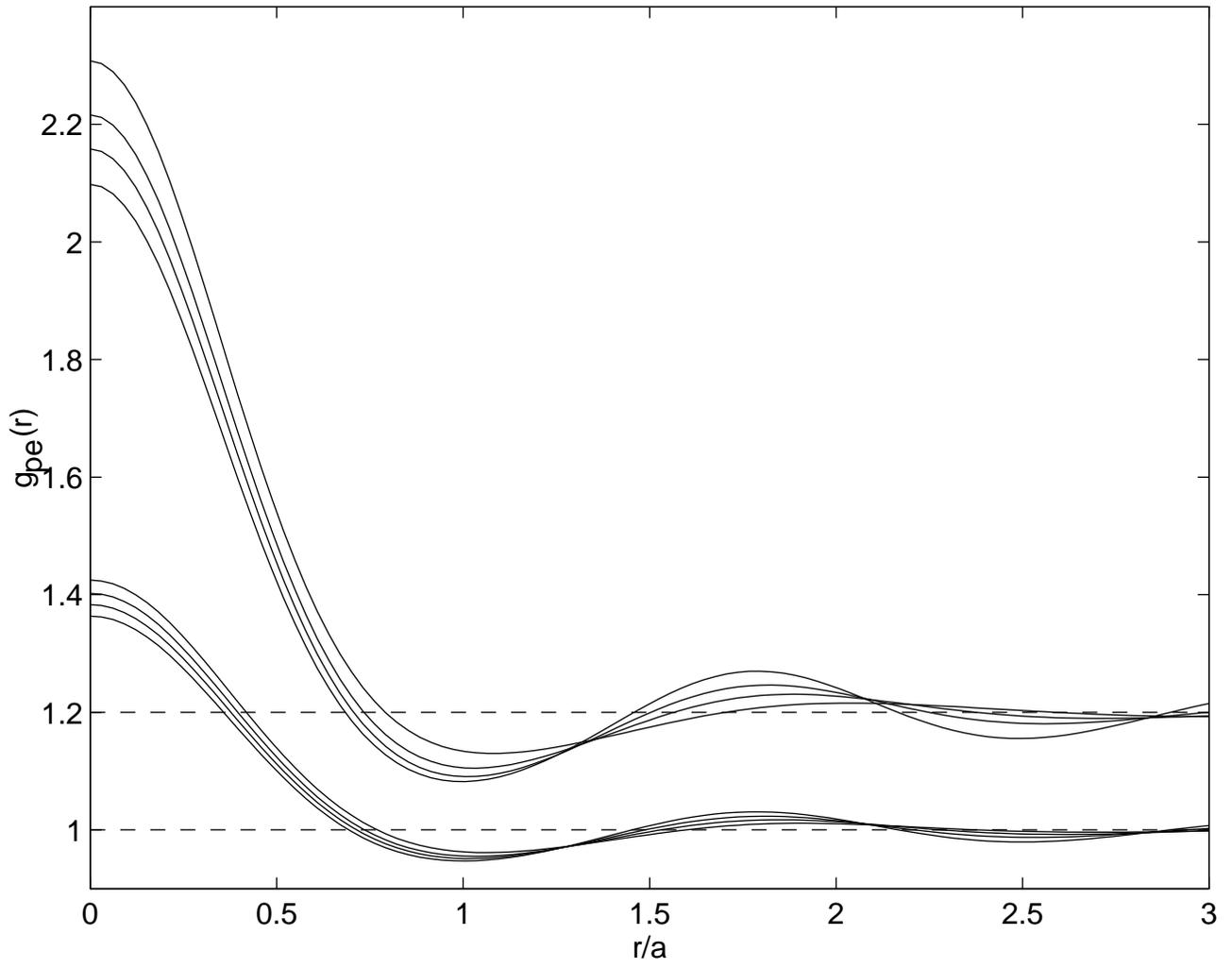
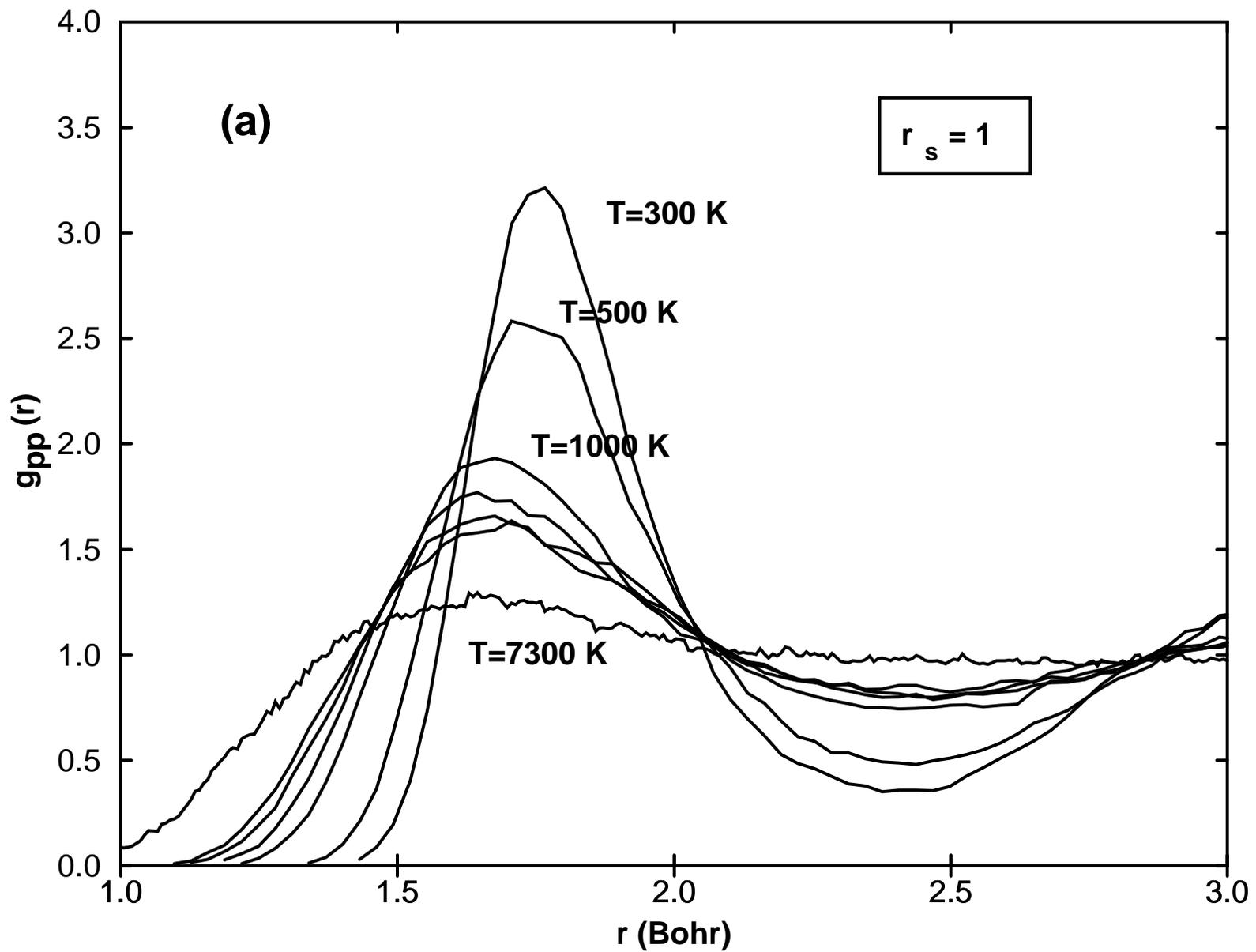
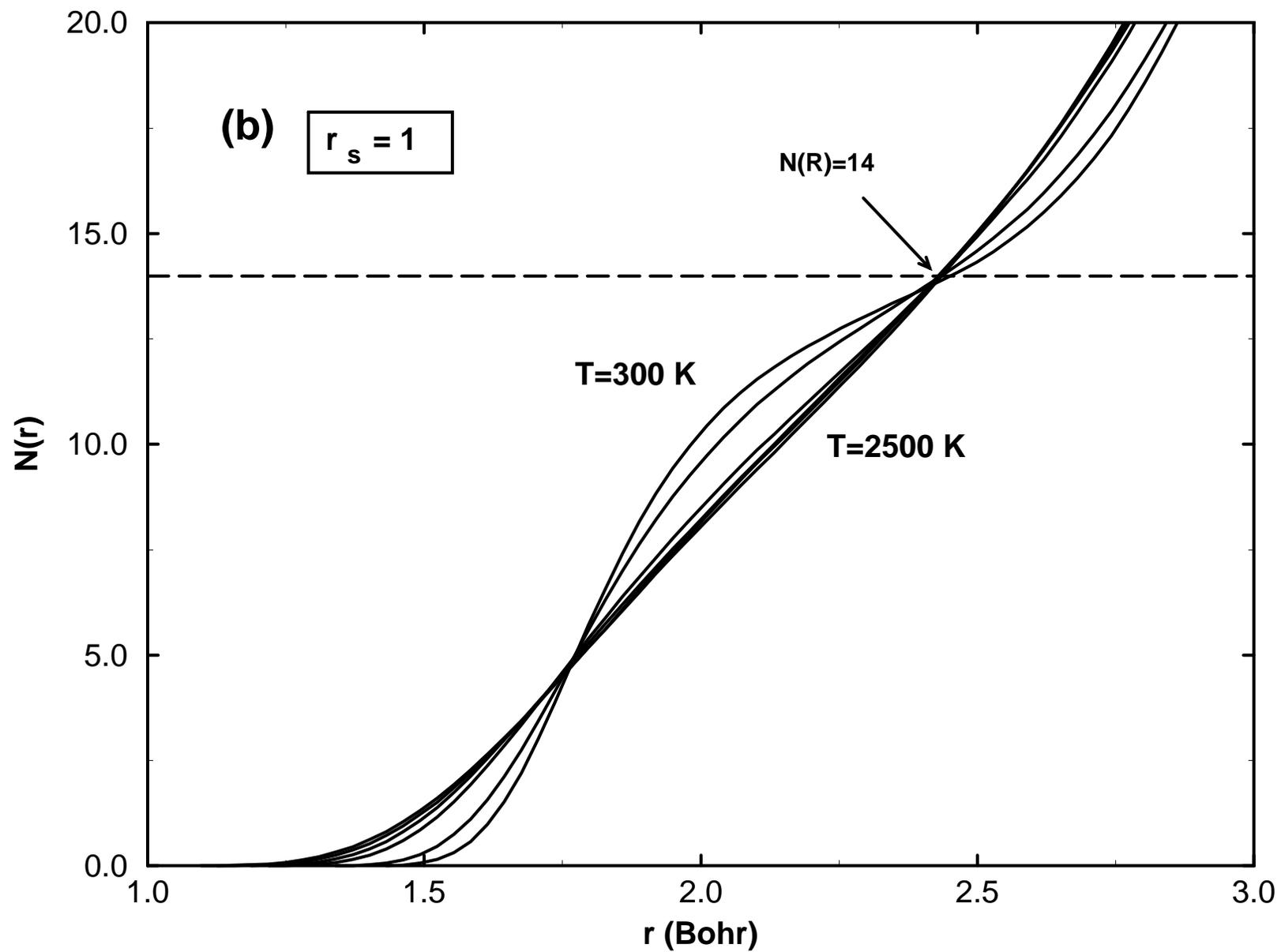
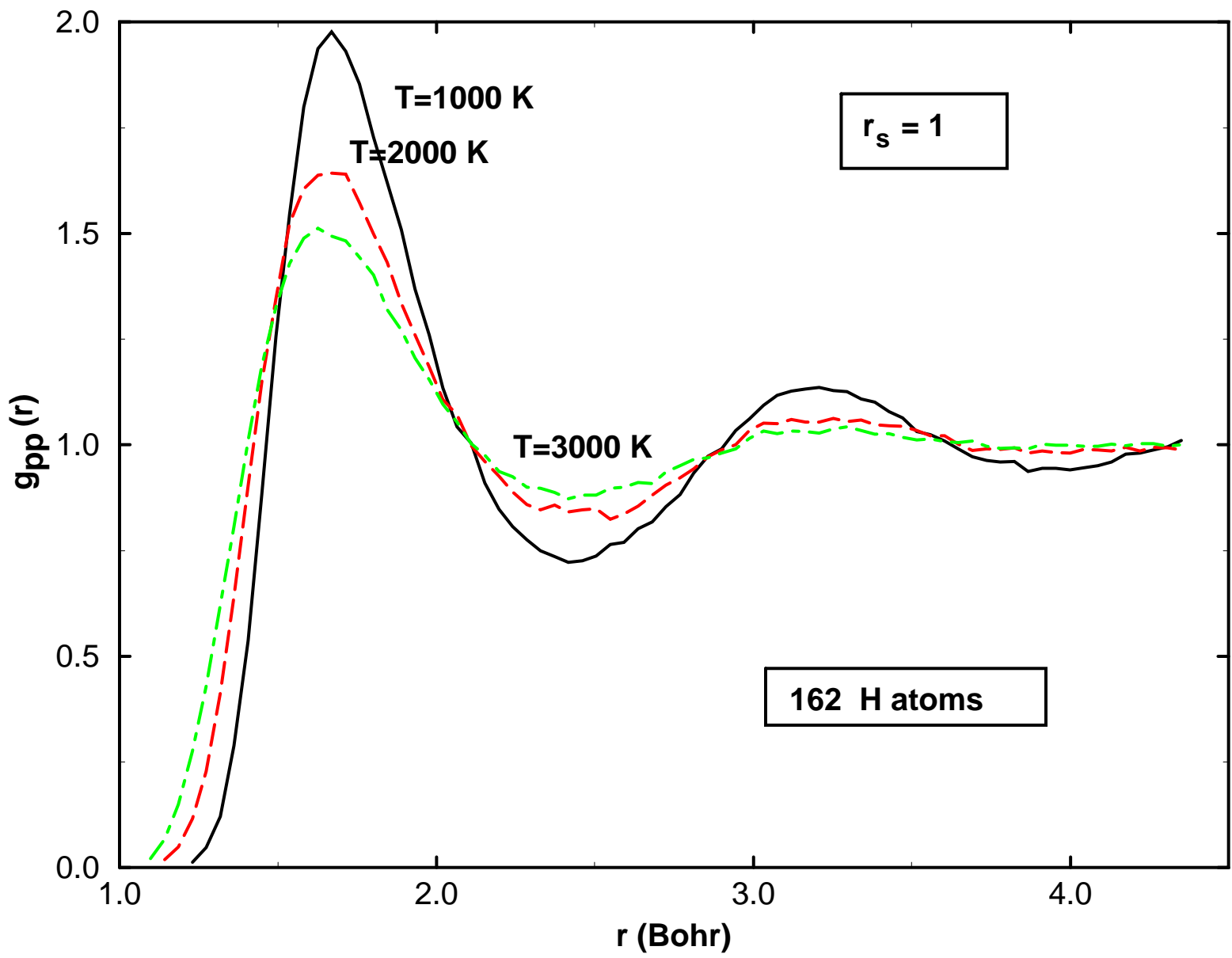
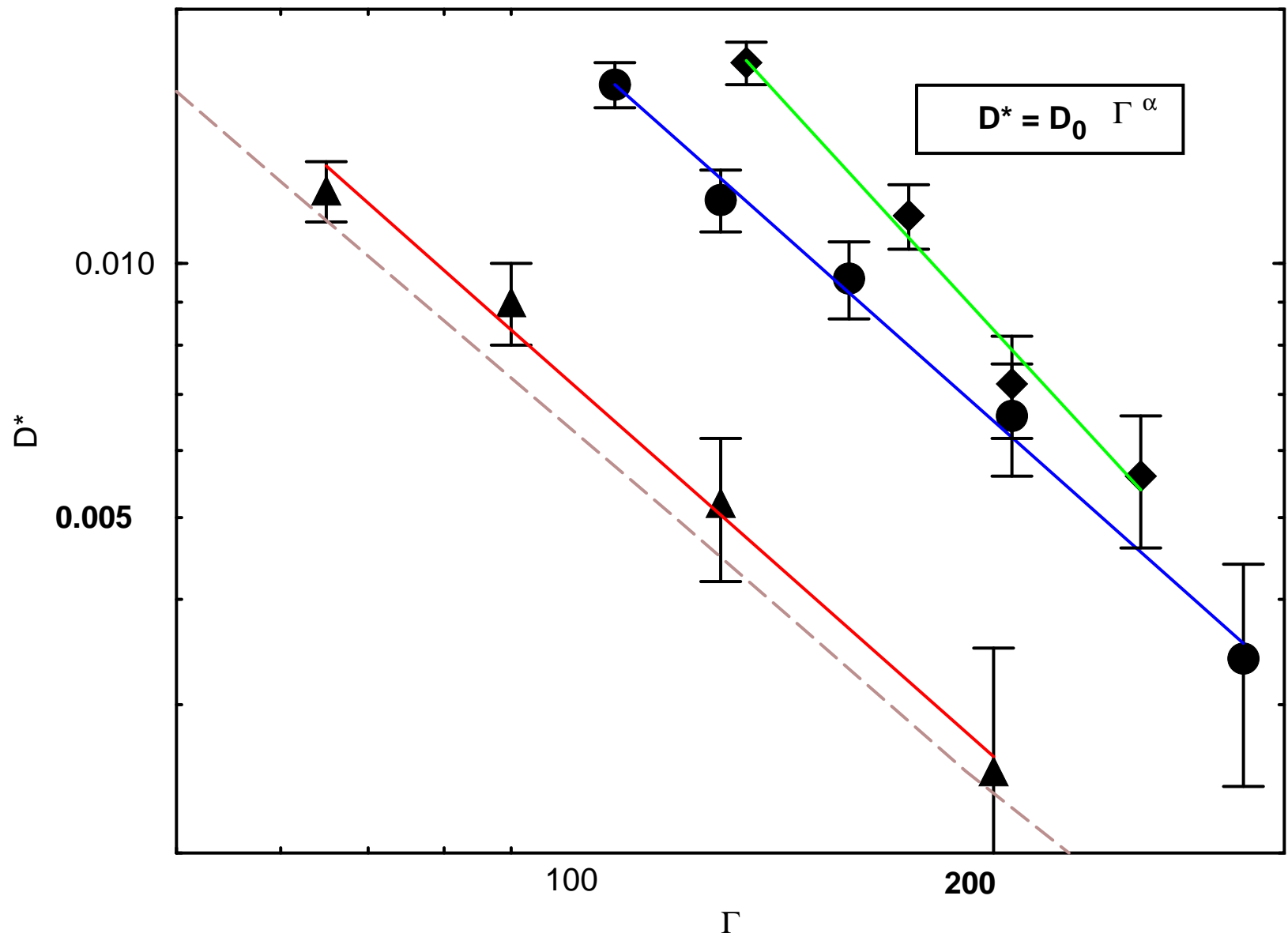
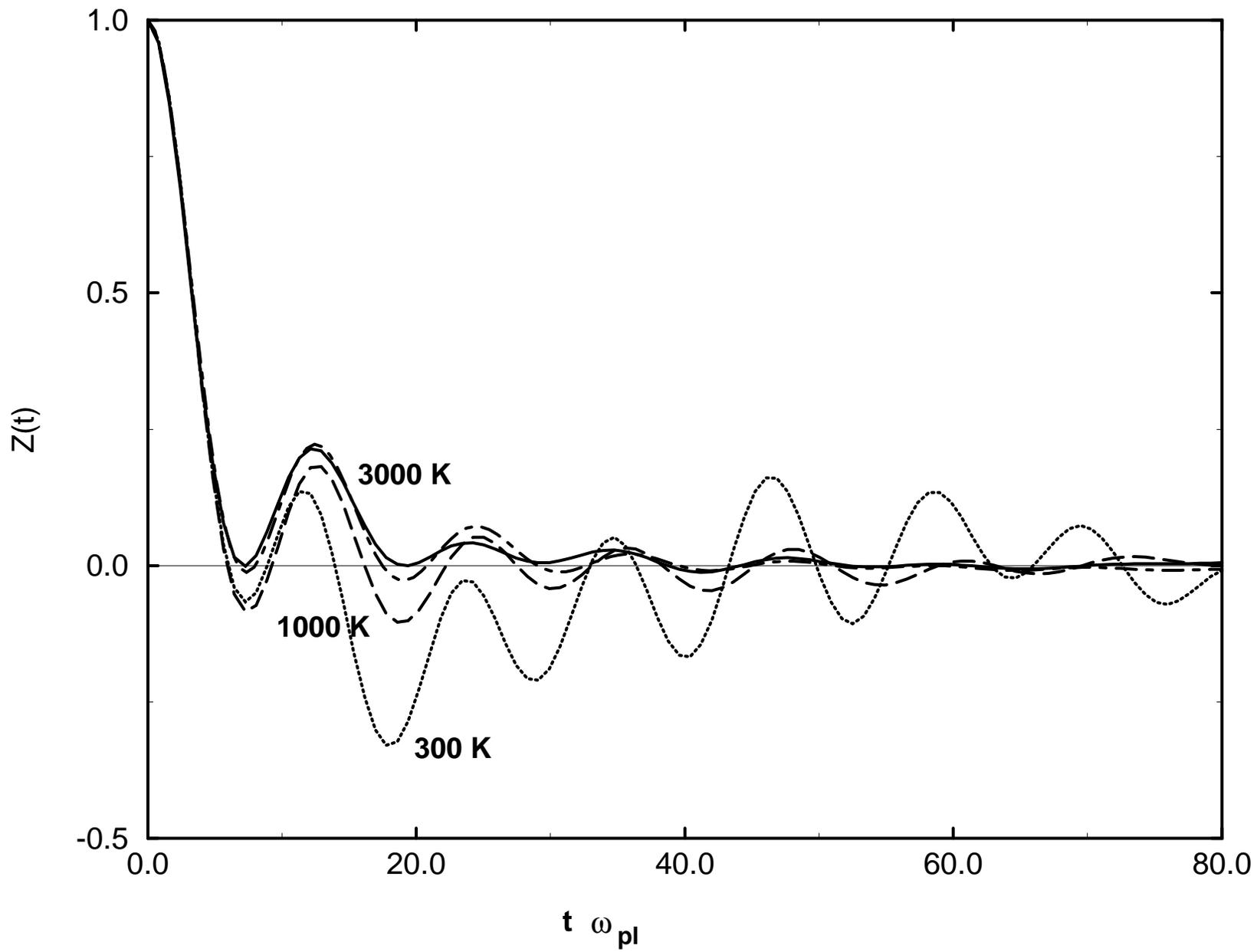
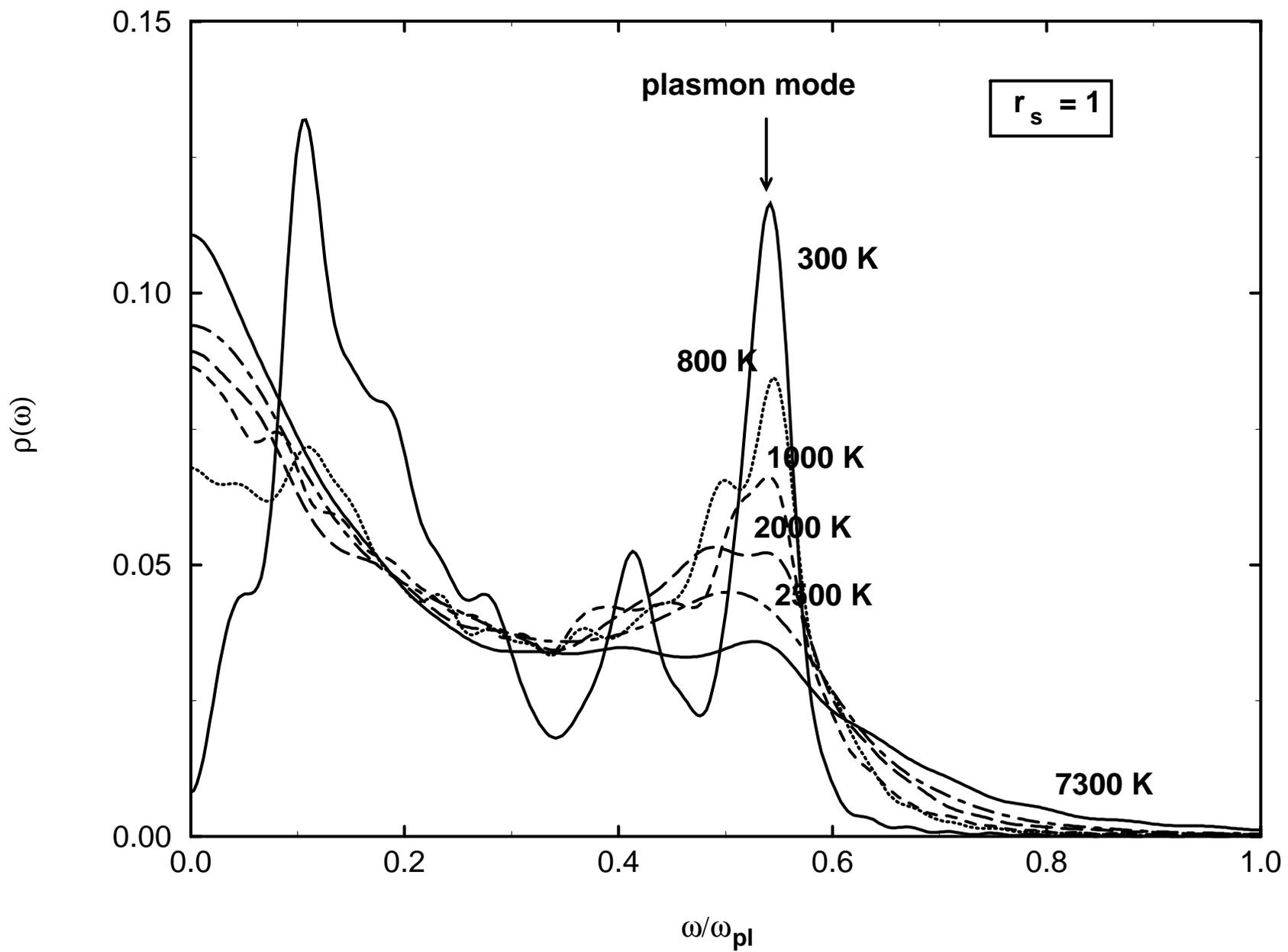
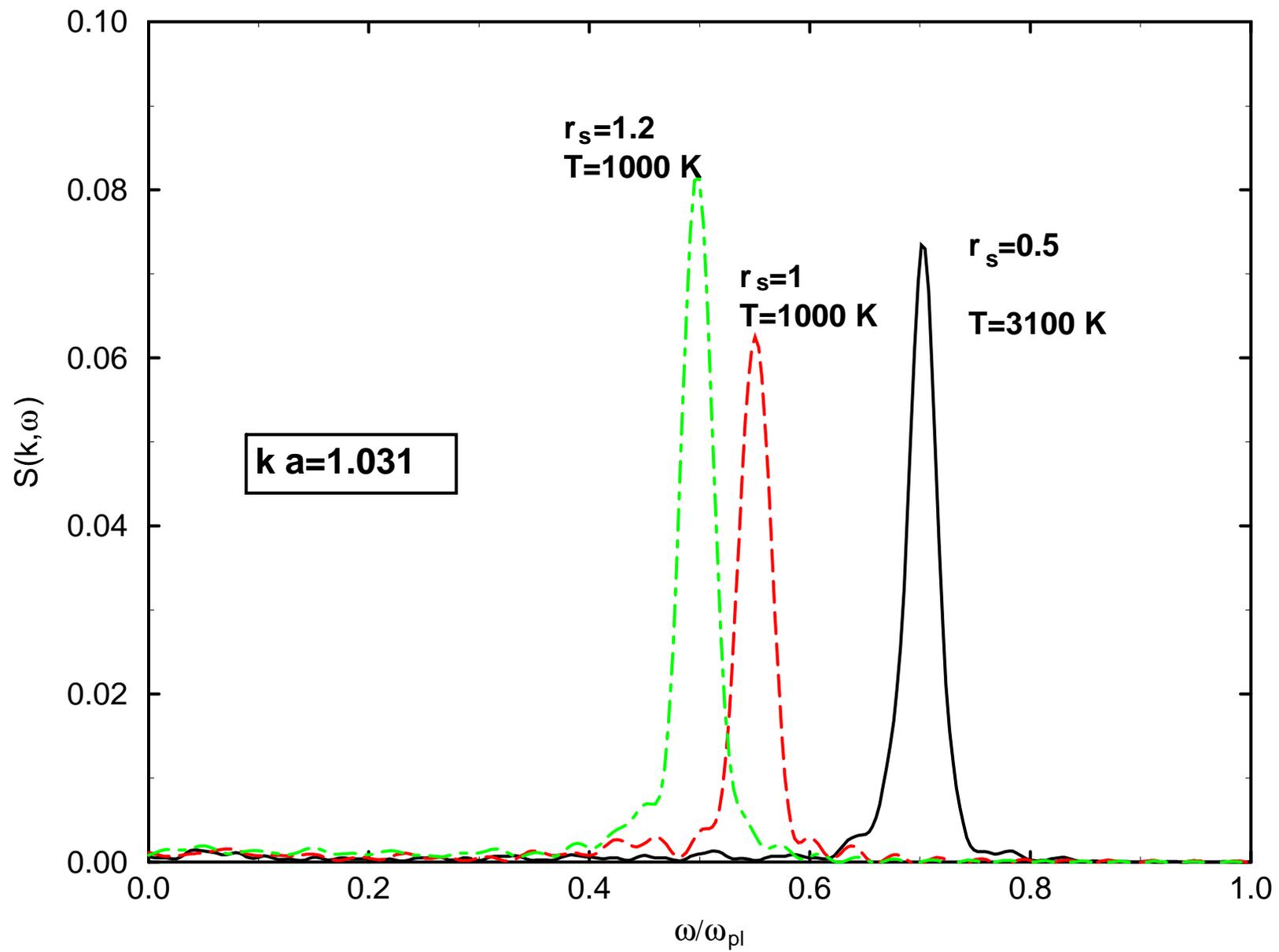
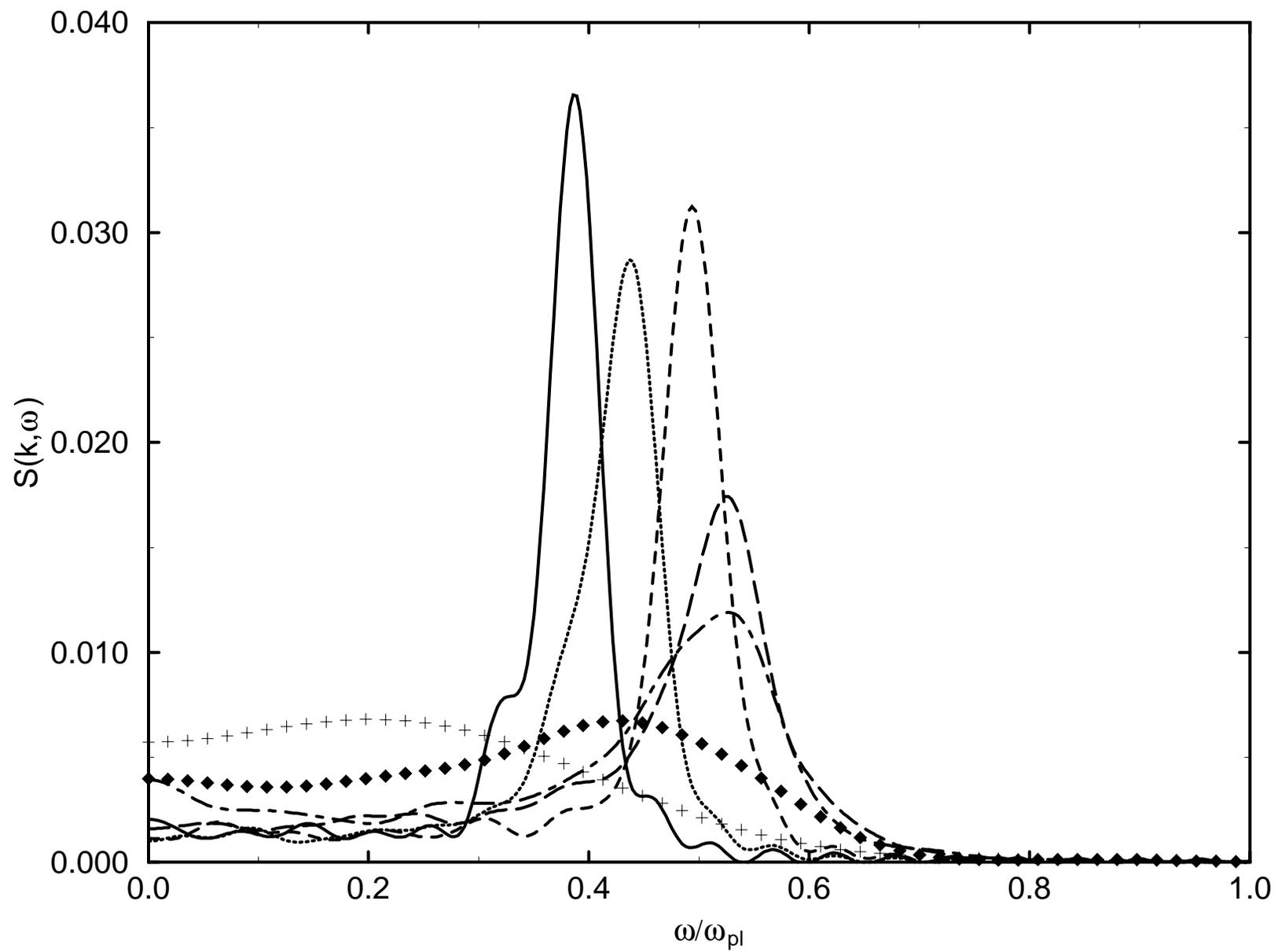
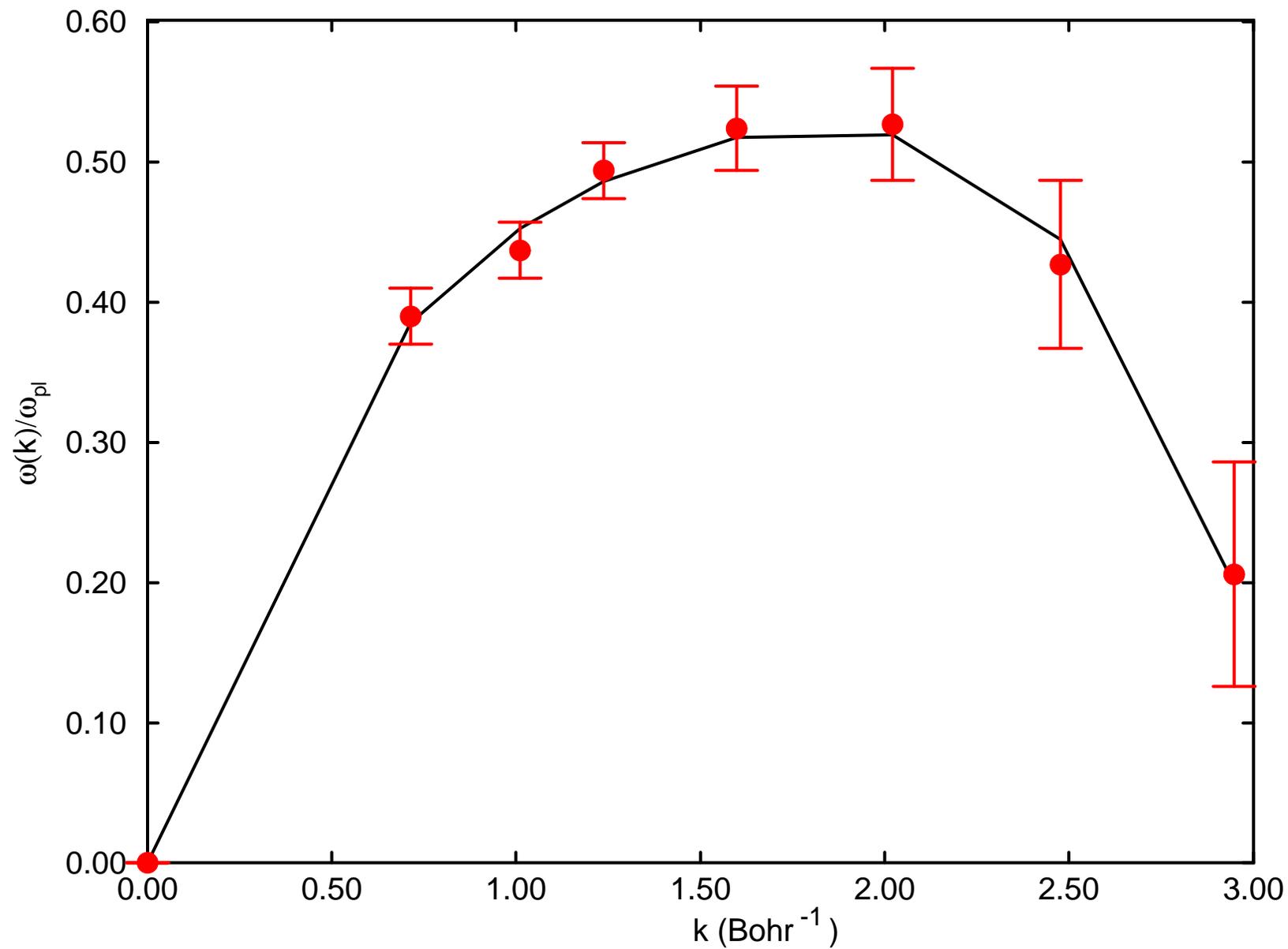
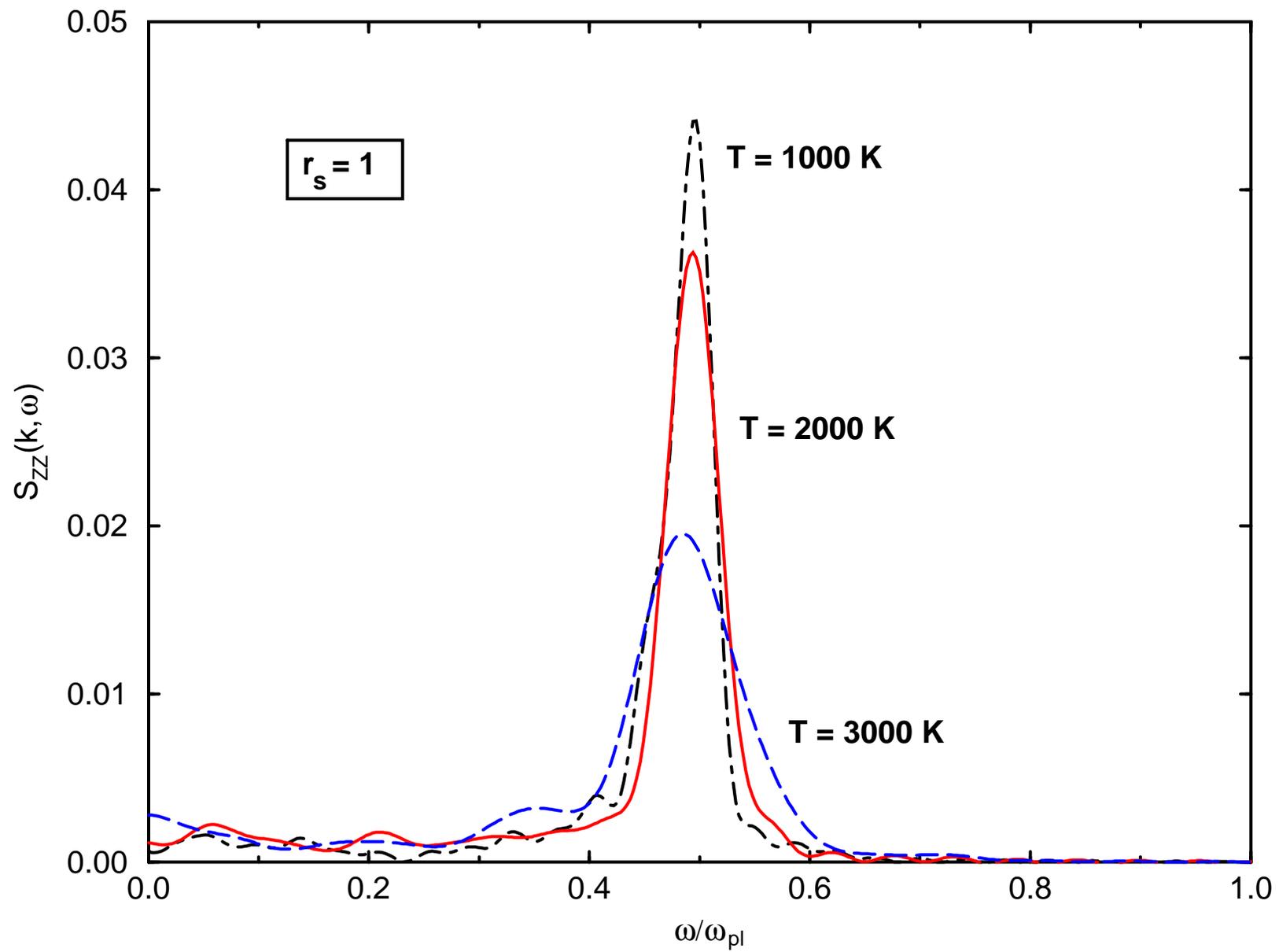
\end{document}